\documentclass[reprint, a4paper, twocolumn,
superscriptaddress,
groupedaddress,
nofootinbib,
 amsmath,amssymb,
prb,
floatfix,
]{revtex4-2}

\usepackage{graphicx}% Include figure files
\usepackage{dcolumn}% Align table columns on decimal point
\usepackage{bm}% bold math
\usepackage{hyperref}
\usepackage{cleveref}
\usepackage[bottom]{footmisc}
\usepackage{natbib}
\usepackage{siunitx} 
\usepackage{xspace}
\usepackage{xcolor}

\newcommand{\ie}{\emph{i.e.}\xspace}
\newcommand{\eg}{\emph{e.g.}\xspace}

\begin{document}

\preprint{APS/123-QED}

\title{Experimental Observation of Short-Range Magnetic Correlations in Amorphous Nb$_2$O$_5$ and Ta$_2$O$_5$ Thin Films}

\author{Y. V. Krasnikova}
\email{yuliak@fnal.gov}
\affiliation{Superconducting Quantum Materials and Systems Center (SQMS),
Fermi National Accelerator Laboratory, Batavia, IL 60510, USA}

\author{A. A. Murthy}
\affiliation{Superconducting Quantum Materials and Systems Center (SQMS),
Fermi National Accelerator Laboratory, Batavia, IL 60510, USA}

\author{D. Bafia}
\affiliation{Superconducting Quantum Materials and Systems Center (SQMS),
Fermi National Accelerator Laboratory, Batavia, IL 60510, USA}

\author{F. Crisa}
\affiliation{Superconducting Quantum Materials and Systems Center (SQMS),
Fermi National Accelerator Laboratory, Batavia, IL 60510, USA}

\author{A. Clairmont}
\affiliation{Superconducting Quantum Materials and Systems Center (SQMS),
Fermi National Accelerator Laboratory, Batavia, IL 60510, USA}

\author{Z. Sung}
\affiliation{Superconducting Quantum Materials and Systems Center (SQMS),
Fermi National Accelerator Laboratory, Batavia, IL 60510, USA}

\author{J. Lee}
\affiliation{Superconducting Quantum Materials and Systems Center (SQMS),
Fermi National Accelerator Laboratory, Batavia, IL 60510, USA}

\author{M.~Shinde}
\affiliation{Superconducting Quantum Materials and Systems Center (SQMS),
Fermi National Accelerator Laboratory, Batavia, IL 60510, USA, Department of Physics, Illinois Institute of Technology, Chicago, IL, 60616, USA}

\author{A.~Cano}
\affiliation{Center for Research and Advanced Studies of the National Polytechnic Institute (Cinvestav), \\Mexico City 07360, Mexico}

\author{D. M. T. van Zanten}
\affiliation{Superconducting Quantum Materials and Systems Center (SQMS),
Fermi National Accelerator Laboratory, Batavia, IL 60510, USA}

\author{M. Bal$^{1}$}
\affiliation{Superconducting Quantum Materials and Systems Center (SQMS),
Fermi National Accelerator Laboratory, Batavia, IL 60510, USA}
\footnotetext{Current affiliation: \textit{Xanadu, Toronto, ON M5B 2H7, Canada}}

\author{A. Romanenko}
\affiliation{Superconducting Quantum Materials and Systems Center (SQMS),
Fermi National Accelerator Laboratory, Batavia, IL 60510, USA}

\author{A. Grassellino}
\affiliation{Superconducting Quantum Materials and Systems Center (SQMS),
Fermi National Accelerator Laboratory, Batavia, IL 60510, USA}

\author{R. Dhundhwal}
\affiliation{IQMT,~Karlsruhe~Institute~of~Technology,~76131~Karlsruhe,~Germany}

\author{D. Fuchs}
\affiliation{IQMT,~Karlsruhe~Institute~of~Technology,~76131~Karlsruhe,~Germany}

\author{T. Reisinger}
\affiliation{IQMT,~Karlsruhe~Institute~of~Technology,~76131~Karlsruhe,~Germany}

\author{I. M. Pop}
\affiliation{IQMT,~Karlsruhe~Institute~of~Technology,~76131~Karlsruhe,~Germany}
\affiliation{PHI,~Karlsruhe~Institute~of~Technology,~76131~Karlsruhe,~Germany}
\affiliation{Physics~Institute~1,~Stuttgart~University,~70569~Stuttgart,~Germany}

\author{A. Suter}
\affiliation{PSI Center for Neutron and Muon Sciences, 5232 Villigen PSI, Switzerland}

\author{T. Prokscha}
\affiliation{PSI Center for Neutron and Muon Sciences, 5232 Villigen PSI, Switzerland}

\author{Z. Salman}
\affiliation{PSI Center for Neutron and Muon Sciences, 5232 Villigen PSI, Switzerland}

\date{\today}% It is always \today, today,

\begin{abstract}

We use muon spin rotation/relaxation/resonance ($\mu$SR) to investigate the magnetic properties of niobium pentoxide (Nb$_2$O$_5$) and tantalum pentoxide (Ta$_2$O$_5$) thin films. In both oxides, we observe a magnetic response at the lowest available temperature of 2.8 K. This response appears to be structurally dependent: thermally oxidized Ta$_2$O$_5$ with low crystallinity demonstrates suppressed magnetism, while fully amorphous Ta$_2$O$_5$ demonstrates local static magnetism. In contrast, amorphous Nb$_2$O$_5$ is dominated by magnetic fluctuations and is strongly magnetically disordered compared to Ta$_2$O$_5$. Our results suggest that these fundamental differences in the magnetism of Ta and Nb oxides could explain the performance limitations in superconducting qubits and resonators.

\end{abstract}

\maketitle

%\tableofcontents

\section{Introduction}

Superconducting qubits are highly sensitive to losses that arise from defects and impurities at material interfaces and surfaces \cite{Krantz2019}. One of the main contributors to these losses are the naturally growing surface oxides that form on the materials used in superconducting quantum devices. In the case of niobium-based devices, niobium pentoxide (Nb$_2$O$_5$) forms natively on the surface, introducing potential sources of microwave losses \cite{Romanenko_Schuster2017, Romanenko2020}. These losses are often attributed to oxygen vacancies within the oxide layer, which can lead to localized magnetic behavior that degrades qubit coherence \cite{Greener1961, Streiff1971, Cava1991, Proslier2011, Wenskat2022, Murthy2022, Bafia2024}. Understanding the mechanisms introducing these losses is critical for improving the performance of superconducting qubits and resonators.

Tantalum-based devices have demonstrated significantly improved coherence time compared to niobium-based devices \cite{Wang2022, Place2021, bland2025millisecond}. Specifically, recent findings by the Superconducting Quantum Materials and Systems (SQMS) group have shown that niobium qubits capped with tantalum metal exhibit up to five times longer coherence times compared to uncapped niobium devices \cite{Bal2024}. This suggests that Ta$_2$O$_5$, the native surface oxide of tantalum, is less lossy than Nb$_2$O$_5$. However, further studies are required to understand the underlying reasons driving this performance improvement. Recent theoretical work Ref.~\cite{pritchard2025suppressed} suggests that these performance differences may arise from the fact that magnetic moments in amorphous Nb$_2$O$_{5-x}$ are far more likely to arise compared to magnetic moments in Ta$_2$O$_{5-x}$. Testing this theory requires studying the local magnetic behavior of Ta$_2$O$_5$ and Nb$_2$O$_5$ using unique capabilities that are highly sensitive to magnetism in amorphous oxide phases.

To this end, we use muon spin spectroscopy ($\mu$SR), a sensitive method to probe magnetic phenomena, to investigate how the intrinsic properties of Nb$_2$O$_5$ and Ta$_2$O$_5$ thin films might contribute to decoherence in superconducting devices. Our primary goal is to determine which of these oxides introduces fewer losses in the microwave range, thereby identifying the material that is less detrimental to qubit performance.

The ultra-thin nature of the surface oxide of Nb and Ta and the superconductivity of the underlying metallic phase, make it difficult to directly investigate them with low-energy $\mu$SR (LE-$\mu$SR). For this reason, we explore the magnetic properties of Nb$_2$O$_5$ and Ta$_2$O$_5$ in thin films prepared by reactive sputter deposition on silicon substrates. We confirm their chemical similarity to the surface oxides using x-ray photoelectron spectroscopy (XPS). In addition, we prepared Ta$_2$O$_5$ films by thermally oxidizing tantalum films deposited by magnetron sputtering on sapphire, to investigate how thermal treatment affects magnetic properties. Throughout this study, we made efforts to reduce the influence of extrinsic factors such as surface contamination and substrate effects, thereby facilitating the evaluation of the intrinsic magnetic behavior of these oxides. Our results present a comparative analysis of the magnetic properties of the two materials, establishing a basis for assessing their potential contributions to loss in superconducting devices.

\section{$\mu$SR spectroscopy}
Muon spin rotation/relaxation/resonance ($\mu$SR) is a technique similar to both electron spin resonance (ESR) and nuclear magnetic resonance (NMR). It enables the measurement of dynamic magnetic phenomena across a frequency range of $10^{3} - 10^{12}$~Hz \cite{Blundell2021}. 
In all experiments described afterwards, we used positive muons, $\mu^+$. The lifetime of a muon is approximately $\tau_\mu=2.2$\;$\mu$s. From the muon decay products:
$$ \mu^+ \to e^+ + \nu_e + \bar{\nu}_\mu
$$
only the positron, $e^+$, is measured. It is preferentially emitted along the direction of the muon’s spin. 
Typically several million $\mu^+$ are implanted into the material to be studied. The recorded time differential positron spectra, originating from a time ensemble of $\mu^+$ decays, of various positron detectors, allow to measure local static and dynamic magnetic fields in solids. A positron spectrum of detector $i$ takes the form:
\begin{equation}
 N_i(t) = N_{i,0} \, e^{-t/\tau_\mu} \; \left[1 + A(t) \right] + N_{i, {\rm bkg}},
\end{equation}
where $N_{i,0}$ gives the scale of measured positrons, $N_{i, {\rm bkg}}$ is a time-independent uncorrelated background, and $A(t)$ is the asymmetry function which is proportional to the muon spin-polarization. Since the detector arrangement is fully symmetric, a common asymmetry $A(t)$ has been used.

For most $\mu$SR studies, the so called surface muons with an energy of about 4\,MeV are used. At this energy, the muon stops at a depth of a couple of 100\,$\mu$m, depending on the density of the material and the technique therefore cannot be applied to characterizing thin films.
At the Paul Scherrer Institute (PSI), a unique spectrometer operates with a beamline of moderated muons, allowing for energy tuning within the keV range \cite{Morenzoni2014, Prokscha_2008}, corresponding to a tunable depth range of $5-300$ nm. This LE-$\mu$SR technique is particularly well-suited for studying thin film samples, as is the case here, offering precise, depth-resolved measurements.

In order to be able to associate a particular depth range with the average energy of the positive muons, we used the TRIM.SP \cite{Eckstein1991} program to calculate the muon stopping profile. The program uses the chemical composition and the material density as input parameters and runs on Monte Carlo simulations. For an energy of 15 keV, the average penetration depth of the muons is about 100 nm for the investigated materials. Stopping profiles for Nb$_2$O$_5$ and Ta$_2$O$_5$ are presented in Sec.\,\ref{sec:appendix_a} Fig.~\ref{fig:profile_Nb2O5}--\ref{fig:profile_Ta2O5}.

The information obtained from the material by $\mu$SR is via the asymmetry, $A(t)$, or more general $A(E,T,B,t)$, where $E$ is the implantation energy, $T$ the temperature, and $B$ the applied magnetic field. The muon is interacting in the material predominantly via magnetic dipolar interaction.
In general there are three different experimental configurations, which will be briefly summarized in the next few sections: 

\begin{enumerate}
    \item In zero field $\mu$SR no external magnetic field is applied, the presence of any magnetic moments in the films can be detected at high sensitivity. In case of an ordered magnetic state, often a coherent precession of the muon spin ensemble can be detected.
    \item In weak transverse field $\mu$SR insights into the magnetic precession of muons in the presence of an external magnetic field and for example any possible dampening due to magnetic ordering, can be gained. Especially, the level of homogeneity of an ordered magnetic state can be determined (magnetic volume fraction). 
    \item Longitudinal field  $\mu$SR studies allow us to explore the dynamics of the local magnetic environment by aligning the magnetic field with the muon spins.
\end{enumerate}

\subsection{Zero Field $\mu$SR, ZF-$\mu$SR}

Assuming that we do not have electronic moments present in the material, \ie only nuclear moments are present, each muon will precess around the total field of the nuclear dipole field distribution at its stopping site. The observed time ensemble of the muon spin polarization, which is proportional to the asymmetry can be well approximated by the Gauss Kubo-Toyabe function \cite{Yaouanc2011}:
\begin{eqnarray}\label{eq:GKT}
 A(t) &=& A_0\, P_{z}^{\text{GKT}}(t) = \nonumber \\
 &=& A_0\, \left[ \frac{1}{3}+\frac{2}{3}\exp{(-1/2\Delta^2t^2)}(1-\Delta^2t^2) \right],
\end{eqnarray}
where $\Delta$ is the 2nd moment of the nuclear dipole field distribution at the muon site, and $A_0$ is the instrumental asymmetry. For LE-$\mu$SR, $A_0$ will depend on the implantation energy. Note: the nuclear moments are considered static on the muon time scale to derive Eq.(\ref{eq:GKT}).

In case electronic moments are present, one has to distinguish between isolated paramagnetic centers and exchanged coupled moments. Typically, $\mu$SR is not sensitive to isolated paramagnetic centers since they are fluctuating at far too high frequencies to be picked up by the muon. On the other hand, exchanged coupled electronic moments that undergo a magnetic transition might lead to a coherent zero-field precession of the muon spin ensemble.

\begin{equation}\label{eq:ZFosc}
\begin{split}
 A(t) = A_0 \alpha\, e^{-\lambda_{\rm S} t} + 
 \\
 +A_0 (1-\alpha) \cos(\gamma_\mu B_{\rm loc} t + \phi)  e^{-\lambda_{\rm F} t},
\end{split}
\end{equation}
where $\alpha=1/3$ in the powder average limit, $\lambda_{\rm S}$ is a slow, and $\lambda_{\rm F}$ a fast depolarization rate. Often $\lambda_{\rm F}$ is so large (static disorder), that only a bi-exponential decay can be observed. In a purely static case $\lambda_{\rm S} = 0$.

\subsection{Weak Transverse Field $\mu$SR, TF-$\mu$SR}

In a weak transverse field (TF) setup, a small magnetic field, $B_{\rm ext}$ is applied perpendicular to the muon spin. In a diamagnetic or paramagnetic sample this leads to a precession signal with an angular frequency of $\omega = \gamma_\mu B_{\rm ext}$, assuming the magnetic shift is negligible. However, if the material is in an ordered magnetic state (\eg, FM or AFM), $B_{\rm ext}$ will break the symmetry in $B$-space, This will result in an asymmetry of the form

\begin{equation}\label{eq:TF}
 A(t) = A(T)\, \cos(\gamma_\mu B_{\rm ext} t + \phi)\, e^{-\lambda t},
\end{equation}
where $A(T)$ will have the properties
\begin{equation}
 A(T) = \left\{ \begin{array}{ll}
               A_0, & T>T_{\rm M} \\
               0    & T\to 0,
              \end{array}
\right.
\end{equation}
where $T_{\rm M}$ is the magnetic transition temperature. Here it is assumed that the whole sample is magnetic at low temperature. More, generally one can define the magnetic volume fraction $f_{\rm M}$ as
\begin{equation}\label{eq:f_M}
 f_{\rm M}(T) = 1 - \frac{A(T)}{A(T>T_{\rm M})}.
\end{equation}
$f_{\rm M}(T)$ is a measure of which fraction of the sample is fully magnetic, \ie $B_{\rm loc} \gg B_{\rm ext}$.

\subsection{Longitudinal Field $\mu$SR, LF-$\mu$SR}

Whereas ZF and TF measurements reveal predominantly static properties of magnetic systems, longitudinal field $\mu$SR is used to obtain information about the dynamic magnetic properties assuming that they fit the time window of $\mu$SR. In a longitudinal field $\mu$SR setup, an external field is applied collinear to the muon spin. This leads to a Zeeman splitting of the muon spin energy levels. Since the involved energies are very small, the muon spin would be locked to its initial state. It can only be flipped being coupled to magnetic fluctuations, which can absorb the corresponding Zeeman energy (2nd order process).

The most widely used model for describing magnetic fluctuations in $\mu$SR is the dynamical Kubo-Toyabe model~\cite{Yaouanc2011}. This is a mean-field approximation that incorporates both static and dynamic components into the depolarization function. In this model, the rate of fluctuations represents the average frequency of ``collisions'' or interactions that a muon experiences due to a stochastically varying magnetic field. Each muon in the sample undergoes depolarization as it encounters these random fluctuations. The hopping rate, denoted by $\nu$, represents the time interval between these collisions, $\Delta$ -- width of the distribution of static local magnetic field, describes the standard deviation of the local magnetic field distribution. For cases of fast fluctuations ($\nu/\Delta \gg 1$), an analytical solution is available \cite{Blundell2021}. However, in the intermediate regime, the asymmetry function can only be solved numerically.

For components of local field:
\begin{equation}
    \overline{B_{\text{loc}}(t_{0})B_{\text{loc}}(t_{0}+t)}=\overline{(B_{\text{loc}})^2}\exp{(-\nu\lvert t \rvert)},
\end{equation}
$\nu$ -- correlation frequency of the random local field $B_{\text{loc}}$ as well as fluctuations rate.

The polarization, and hence the asymmetry $A(t) = A_0\, P_{z}(t)$, could be defined in case of dynamic input presence by integral equation:
\begin{equation}\label{eq:dynGKT_LF}
\begin{split}
P_{z}(t)=P_{z}^{\text{stat}}(t)\exp({-\nu t})+\\
+\nu \int_{0}^{t}P_{z}(t-t'){P_{z}^{\text{stat}}(t')\exp({-\nu t'})dt'},
\end{split}
\end{equation}
In case when $\nu=0$ polarization is defined by static Kubo-Toyabe function in an applied field $B_{\rm ext}$:
\begin{equation}
\begin{split}
P_{z}^{\text{stat}}(t)= 1 - \frac{2\Delta^2}{\omega^2}\, \left[ 1 - e^{-(\Delta\cdot t)^2/2} \cos(\omega t) \right] +\\
  + \frac{2\Delta^4}{\omega^3}\, \int_0^t e^{-(\Delta\cdot \tau)^2/2} \sin(\omega \tau)\, d\tau,
\end{split}
\end{equation}
where $\omega = \gamma_\mu B_{\rm ext}$, and $\Delta$ is the 2nd moment of the static field distribution. For $B_{\rm ext}=0$ (ZF case), this reduces to Eq.(\ref{eq:GKT}).

\section{Experimental results}

In the next section, we outline the sample preparation and characterization in detail.

\subsection{Sample characterization}

Nb$_2$O$_5$ films were prepared using reactive sputter deposition. A Si wafer is loaded inside an AJA ATC 2200 sputtering system with a base pressure better than $10^{-8}$ Torr. DC magnetron sputtering was performed using a 3-inch diameter Nb target with a metals basis purity of 99.95\% with an Ar flow rate of 30 sccm, an O$_2$ flow rate of 20 sccm, and partial pressure of 3.5 mTorr at room temperature. A sputtering power of 600 W was used and the substrate was rotated at 20 rpm. These conditions resulted in a Nb$_2$O$_5$ deposition rate of approximately 5 nm/min. Si wafers were diced in pieces of size $20$ mm x $20$ mm, a few samples were used for muon spectroscopy, others for materials characterization. 

The sputtered Ta$_2$O$_5$ films were prepared using a 3-inch diameter tantalum pentoxide target in an AJA ATC Orion 8 UHV sputtering system. RF magnetron sputtering with a power 400 W was used with an Ar flow rate of 27 sccm and an O$_2$ flow rate of 3 sccm at a partial pressure of 5 mTorr at room temperature. The sputtered Ta$_2$O$_5$ film was deposited at a rate of 7 nm/min on a Si wafer, where the substrate was rotated at 20 rpm.

For the thermal Ta$_2$O$_5$ films we used a DC magnetron sputter deposition tool with base pressure below \SI{1e-8}{\milli\bar}, to deposit epitaxial tantalum films on 2" epi-polished c-plane sapphire from a high purity tantalum target (MSE Supplies, Batch $33222A9$, \SI{99.999}{\percent}). We heat the substrate to a temperature of about \SI{530}{\celsius} during deposition, while rotating it. Substrates are cleaned with Isopropanol (IPA) for $\SI{10}{\minute}$ in ultrasonic bath, and then heated to about \SI{300}{\celsius} in the deposition chamber overnight before deposition. The deposition parameters are summarized in Table I. 
After deposition, films are cooled down at a rate of $\SI{2}{\degreeCelsius\min^{-1}}$. With these deposition parameters the Ta film has predominantly
epitaxial alignment $(0001)$ Al$_{2}$O$_{3}\,\parallel\,(111)\,$Ta with the sapphire substrate. 
We then diced the wafer into smaller pieces, in particular four $15$ mm x $15$ mm to be used in the muon spectroscopy. Before dicing we coated the wafer with protective resist, which we stripped afterwards using several solvent cleaning steps.
In order to oxidize the metallic film we annealed the wafer pieces at
 \SI{500}{\celsius} in air for 24 hours, turning the samples fully transparent with a slight greenish hew. The film thickness increased by about  \SI{25}{\percent} in the oxidation step, resulting in the film thicknesses noted in Table I.  The temperature was chosen to yield a sufficiently high oxidation rate and avoid crystallization of the oxide~\cite{Chandrasekharan2005Dec}. However, X-ray diffraction and atomic-force microscopy of the films revealed that the thick thermal  Ta$_2$O$_5$ film (Sample 1, KIT) is at least partially crystalline, probably to the high degree of order in the precursor Ta film (see Appendix B). 

\begin{table}
\label{table:sputtering_parameters} 
\begin{ruledtabular}
\begin{tabular}{ m{2.8cm} p{1.9cm}  p{1.9cm} p{1.9cm}}
    Parameter & sputtered Nb$_2$O$_5$ & sputtered Ta$_2$O$_5$ & thermal Ta$_2$O$_5$\\
\hline
    Sputter Target & 3" Nb & 3" Ta$_2$O$_5$ & 3" Ta \\
    %Target Purity & \SI{99.95}{\percent} & & \SI{99.999}{\percent} \\ 
    Sample Temp. ($^\circ$C) & 20 & 20 &  530   \\ 
    Ar flow (sccm) & 30 & 27 &  44  \\  
    O$_2$ flow (sccm) & 20 & 3 &  0  \\  
    Pressure (mbar) & 0.005 & 0.007 & 0.003 \\ 
    Power ($\si{\watt}$) & 600 & 400 &  200  \\ 
    Rate (nm/s) & 0.08 & 0.12 & 0.06  \\ 
    Substrate & Si & Si & Sapphire \\
    Film thickness (nm) & 110& 100 & 170 (S1),\newline 105 (S2) \\
    Deposition site & SQMS & SQMS & KIT \\ 
\end{tabular}
\end{ruledtabular}
\caption{Samples' parameters. }
\end{table}

In the next two sections we present the $\mu$SR results characterizing the magnetic properties of the thin films. The $\mu$SR data we analyzed using the \textit{musrfit} software \cite{Suter2012musrfit}. The temperature was set in the range from room temperature down to base temperature 2.8~K using a cold finger $^4$He cryostat. The magnetic field did not exceed 130~mT in the LF experiment and was in the range of  5--7.5~mT in the TF experiments.

\subsection{Nb$_2$O$_5$ -- $\mu$SR Results}
\subsubsection{Weak transverse field results}

\begin{figure}[!htb]
\includegraphics[width=0.5\textwidth]{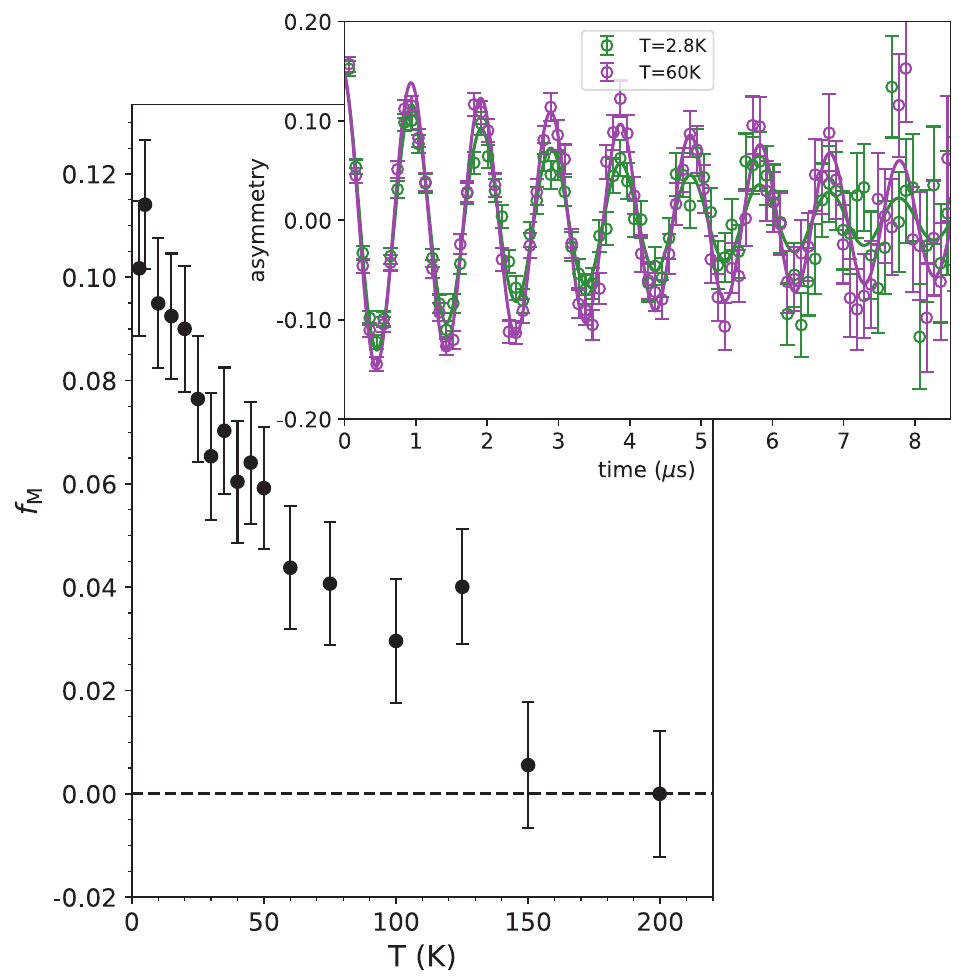}% Here is how to import EPS art
\caption{$\mu$SR data in a transverse magnetic field of  $7.5$~mT  (TF) in Nb$_2$O$_5$, $E=9$~keV (55~nm). Temperature dependence of magnetic volume fraction ($f_M$). inset: Example of TF data for two different temperatures 2.8~K (green) and 60~K (purple). Solid lines are fitting.}\label{fig:fM_Nb2O5}
\end{figure}

In the TF measurements, we applied an external field of $B_{\rm ext}=7.5$ mT. The inset of Fig.~\ref{fig:fM_Nb2O5} shows asymmetry spectra at two different temperatures. As can be seen, there is a slight reduction in the asymmetry between these two data sets. The resulting magnetic volume fraction, $f_M$, shown in  Fig.~\ref{fig:fM_Nb2O5} as a function of temperature, shows a gradual increase, below $T\sim 120$ K, toward lower temperature, reaching a value of about 0.11 at the lowest measured temperature. This means that only a marginal part of Nb$_2$O$_5$ shows signs of static magnetism.

\subsubsection{Zero field results}

\begin{figure}[h]
\includegraphics[width=0.45\textwidth]{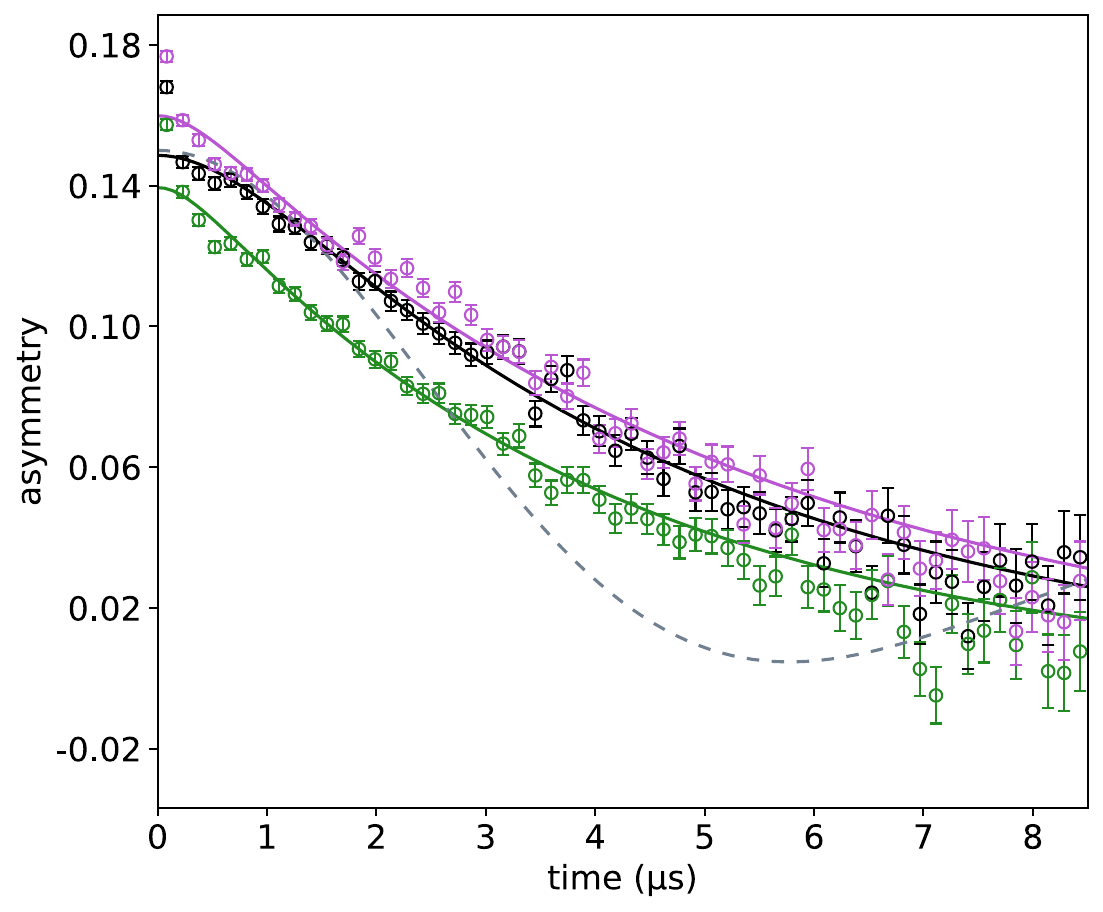}
\caption{Zero field $\mu$SR time spectra in Nb$_2$O$_5$, for different implantation energies (penetration depths $z$) $E= 3, 9, 18$~keV (bottom to top, corresponding to $z \simeq 20, 55, 115$~nm) at base temperature $T=2.8$~K. Dashed line corresponds to  Eq.(\ref{eq:GKT})}\label{fig:ZF_Nb2O5}
\end{figure}

In Fig.~\ref{fig:ZF_Nb2O5} time spectra at different implantation energies are presented. In the absence of any electronic moments, the time spectrum should follow Eq.(\ref{eq:GKT}), due to the nuclear damping ($^{93}$Nb, natural abundance 100\%, $I=9/2$, $\mu/\mu_{\rm N} = 6.1705$). This is depicted by the dashed line in Fig.~\ref{fig:ZF_Nb2O5}.
The time spectra, are obviously not following this shape. The reason is not only the small percentage of static magnetic moments, as revealed by the TF measurements, but also the presence of magnetic fluctuations. The lines in Fig.~\ref{fig:ZF_Nb2O5} are fits to Eq.(\ref{eq:dynGKT_LF}) for $\omega=0$. The data shows that these magnetic fluctuations are present throughout the film.
Note: the upturn in the asymmetry at very early times ($t\lesssim 70$ ns) is due to time-of-flight decay of muons \cite{Suter2023}. Due to the binning this seemingly extends to larger times.

\subsubsection{Longitudinal field results}
The ZF data already show that magnetic fluctuations are present at the lowest measured temperature $T=2.8$~K. However, to quantify the fitting parameters, and hence the magnetic fluctuation rate, it is necessary to perform LF measurements. The reason is that $\Delta$ and $\nu$ (see Eq.(\ref{eq:dynGKT_LF})) are strongly correlated, and only combined measurements at various $B$-fields allow a global fit to reliably obtain the fluctuation rate $\nu$.

\begin{figure}[!htb]
\includegraphics[width=0.48\textwidth]{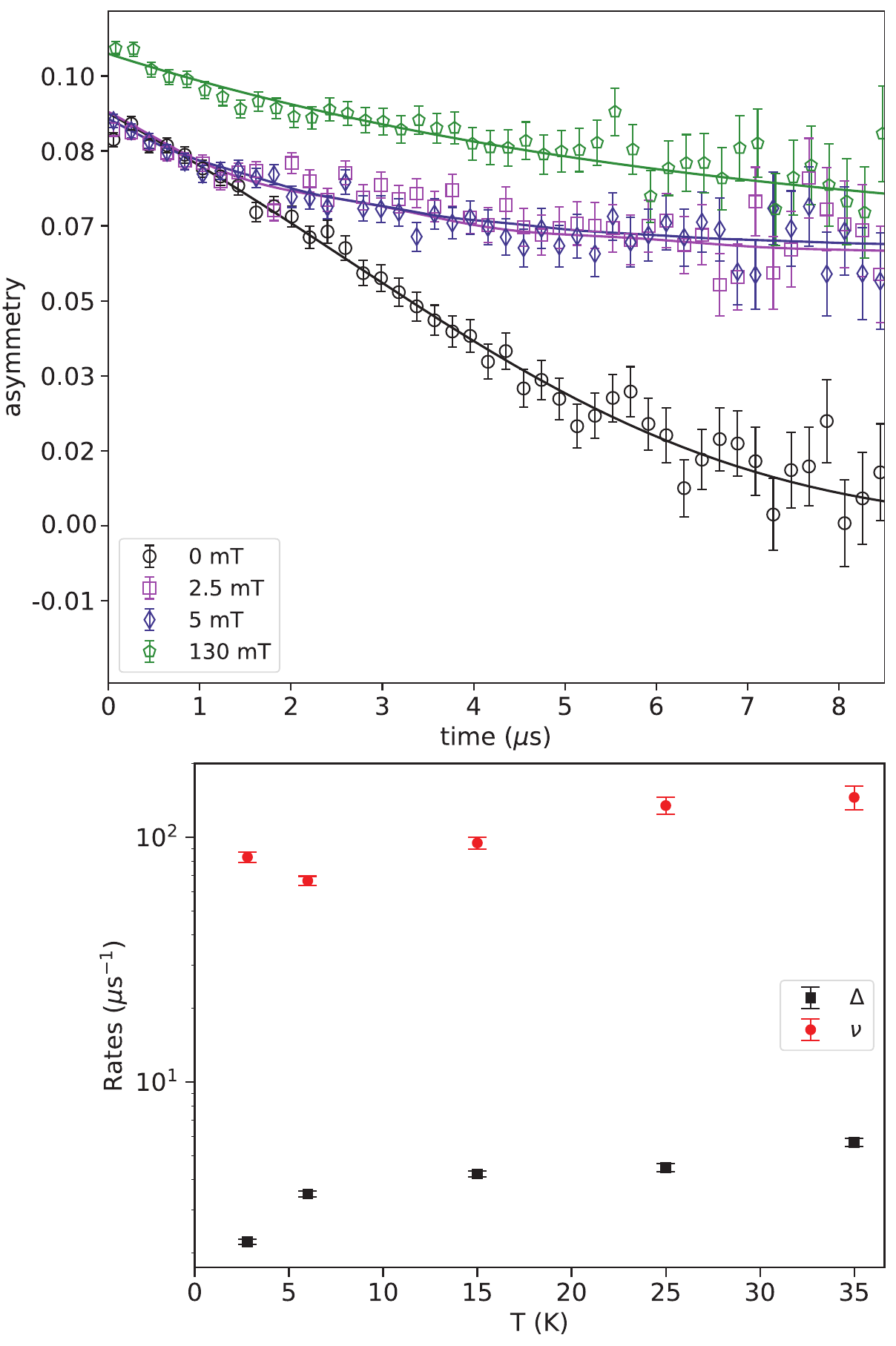}
\caption{\label{fig:LF_NbOx} $\mu$SR data in longitudinal magnetic field (LF) in Nb$_2$O$_5$, $E=9$~keV ($z \simeq 55$~nm), $T=15$~K.  Upper panel: Asymmetry function in different magnetic fields at fixed temperature. Lower panel: results of global fitting for asymmetry function at different temperatures, taking dynamic Kubo-Toyabe function with $\Delta$ and $\nu$ parameters (described in the text).}
\end{figure}

To analyze and fit our data, we used a two-subsystem model. At high temperatures, we connected the behavior of the asymmetry predominantly with the nuclear subsystem. In contrast, at lower temperatures, we assumed that the asymmetry arises from a combination of contributions from both the nuclear and electronic subsystems. The nulcear part is described by Eq.(\ref{eq:dynGKT_LF}) assuming a quasi-static state, \ie $\nu = 0$, and fixed $\Delta_{\rm n}$, whereas the electronic part is described by Eq.(\ref{eq:dynGKT_LF}), with free $(\nu,\Delta)$ as shown in Fig.~\ref{fig:LF_NbOx}. The asymmetries have been chosen to fit the magnetic volume fraction: $A_{\rm n} = A(T) (1-f_{\rm M}(T))$ and $A_{\rm e} = A(T) f_{\rm M}(T)$.  The numerical fitting results of the global fit for niobium pentoxide are shown as solid lines in upper panel of Fig.~\ref{fig:LF_NbOx}. Bottom panel of Fig.~\ref{fig:LF_NbOx} is result of electronic component to the fitting.

In our experiment, the muons are sensitive to fluctuation rates on the order of 100 MHz in Nb$_2$O$_5$, although this can potentially be extended by an order of magnitude towards higher and lower frequencies. The precise width of the frequency distribution is unknown from the experiment.  A reader could imagine it as one data point on the noise spectra curve. Taking a different probe instead of muon would potentially give more data. The source of these fluctuations originates from correlated electronic moments.

\subsection{Ta$_2$O$_5$ -- $\mu$SR Results}
\subsubsection{Weak transverse field results}

%\onecolumngrid
\begin{figure*}[!htb]
\includegraphics[width=1\textwidth]{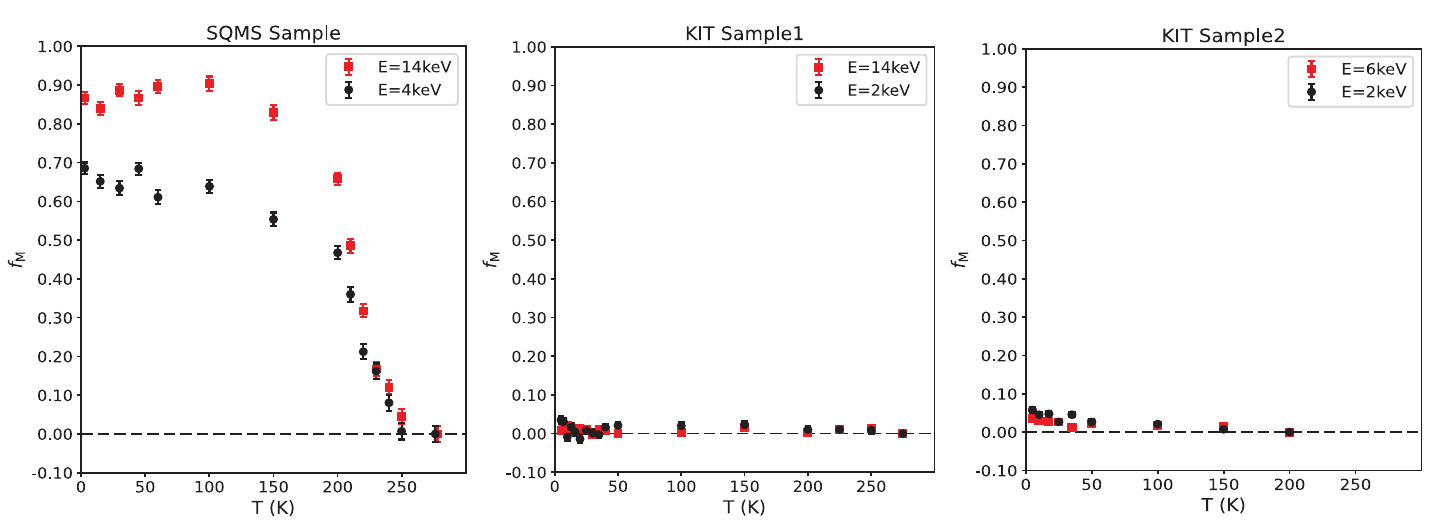}% Here is how to import EPS art
\caption{Temperature dependence of the magnetic volume fraction ($f_M$) in Ta$_2$O$_5$, measured at $B_{\rm ext}=5$~mT. Left panel: data for sputtered oxide film, where black circles show $E=4$~keV (25~nm), and red squares: are $E=14$~keV (75~nm). 
Middle panel: data for thermal oxide films, where black circles show $E=2$~keV (15~nm), and red squares $E=14$~keV (75~nm). Right panel: data for thermal oxide films, where black circles show $E=2$~keV (15~nm), and red squares $E=6$~keV (35~nm).}\label{fig:Ta_oxides_fM}
\end{figure*}
\twocolumngrid

The TF results of the sputtered and thermally grown oxide films show a very different behavior, as can be see in Fig.~\ref{fig:Ta_oxides_fM}. The sputtered films show a rather sharp magnetic transition at $T_{\rm M}=225$~K (defined at the half-way increase point of $f_M$), with a large magnetic volume fraction of $f_M(E=14\,\mathrm{keV})=0.88(2)$, $f_M(E=9\,\mathrm{keV})=0.85(2)$ (not shown), and $f_M(E=4\,\mathrm{keV})=0.64(2)$. On the other hand, the thermally grown films show essentially no magnetic volume fraction $f_M < 0.04(2)$.

\subsubsection{Zero field results}

As the TF results have already shown, the sputtered and thermal grown samples show very different behavior. This is even more pronounced for the ZF results. 

\paragraph{Sputtered Sample}

\begin{figure*}[!htb]
\includegraphics[width=0.9\textwidth]{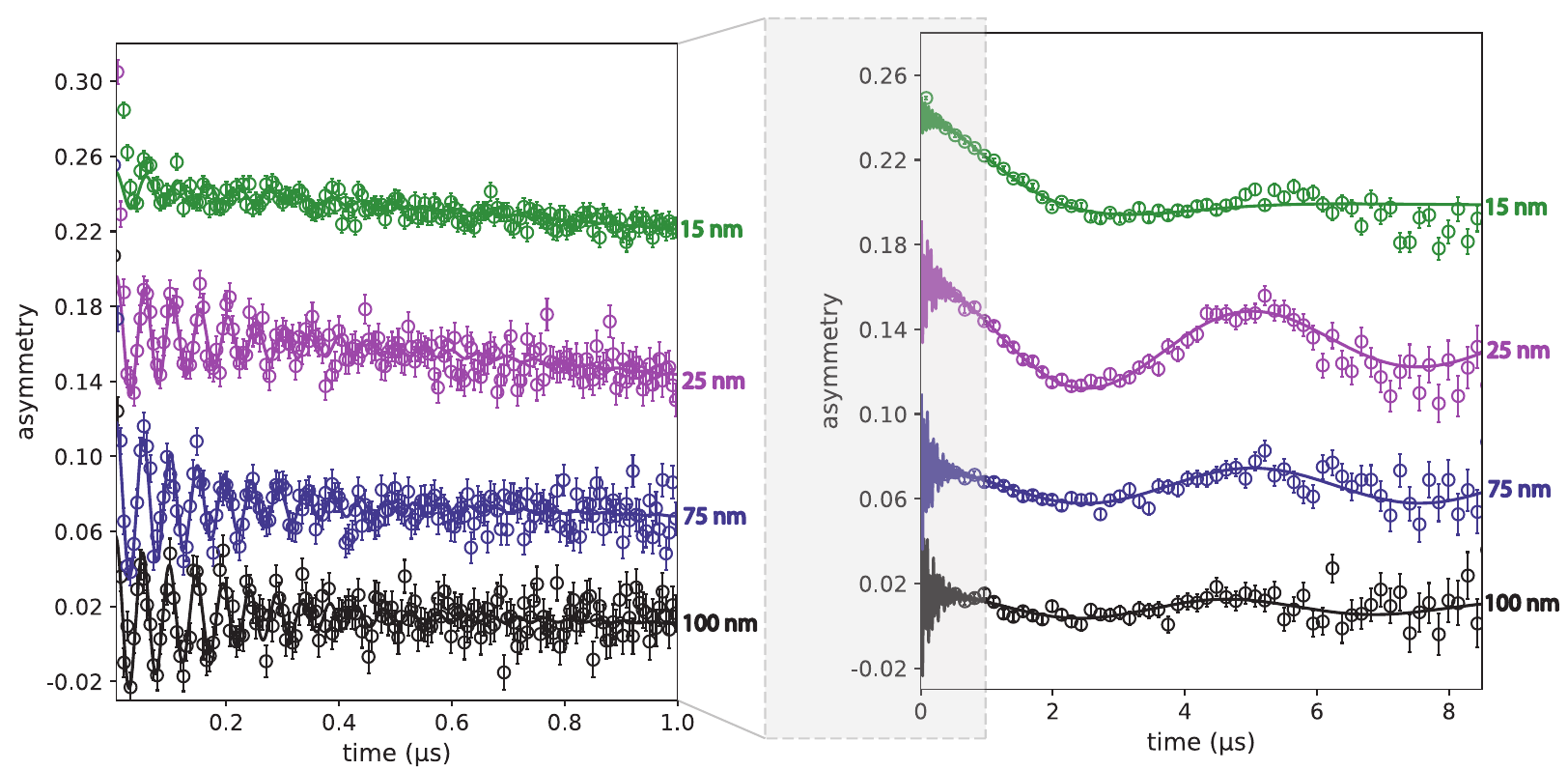}
\caption{Short time (left panel) and long time (right panel) zero field $\mu$SR spectra of sputtered Ta$_2$O$_5$, for different energies (penetration depths): $E= 2, 4, 14, 19$~keV (15, 25, 75, 100~nm) and base temperature $T=2.8$~K. The plots are shifted with respect to each other by 0.06. Grey area represents short time scale.}\label{fig:Ta2O5_sputtered_ZF_both}
\end{figure*}

\begin{figure}[!htb]
 \includegraphics[width=0.4\textwidth]{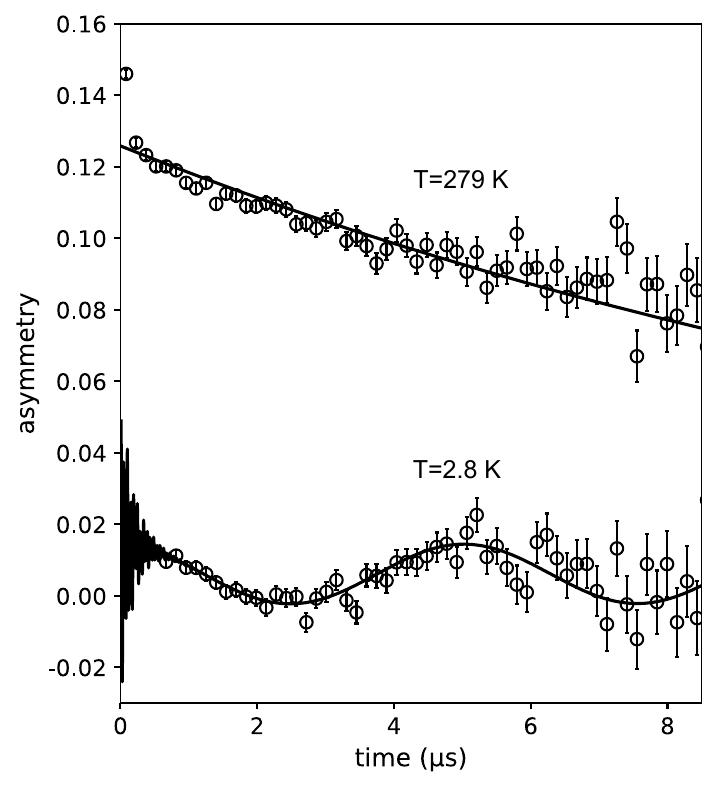}%
 \caption{Zero field $\mu$SR time spectra for sputtered Ta$_2$O$_5$ for $T=2.8$~K and $T=279$~ K, measured for an implantation energy of $E=14$~keV ($z=75$~nm).}\label{fig:Ta2O5_sputtered_ZF_low_and_highT}
\end{figure}

The sputtered sample shows clear zero-field oscillations at the lowest measured temperature. Since there are two distinct frequencies: high and low, they are shown in Fig.~\ref{fig:Ta2O5_sputtered_ZF_both}. Such ZF oscillations are often a signature of an ordered magnetic ground state, in our case likely antiferromagnetically coupled moments present as the transition is not detectable in SQUID magnetometry (see Appendix B), we don't have a conventional long-range order in the system, however clearly strongly correlated magnetic moments are leading to muons' precession. The high frequency corresponds to an internal field of $B_{{\rm loc}, 1} \sim 150$~mT at the muon site, whereas the low frequency is $B_{{\rm loc}, 2} \sim 1.5$~mT. This means that there are two distinct muon sites present in the structure most likely due to two types of defects (caused by oxygen vacancies). Since the coupling of the ordered electronic moment to the muon spin is of magnetic dipolar character, we expect a dependence of the internal field on distance $r$ between the muon and the ordered magnetic moment of $1/r^3$. We can therefore deduce from the ratio of the estimated internal fields, that the low frequency signal is from muons which are about $4.6$ times further away from the ordered electronic moment.  It is also apparent that there is substantially more disorder towards the sample surface since the oscillation amplitude is much more strongly damped. These ZF oscillations have already been washed out at $T=4$~K. Note: It looks as if there are two transitions present: one at around $T_{\rm M,h}=225$~K as deduced from the TF measurements (left panel of Fig.~\ref{fig:Ta_oxides_fM}), and a second one at around $T_{\rm M,l}=3$~K that can be deduced from the ZF oscillations vanishing at $T=4$~K (data not shown). A comparison of ZF time spectra of $T=2.8$~K and $T=279$~K for $E=14$~keV are shown in Fig.~\ref{fig:Ta2O5_sputtered_ZF_low_and_highT}.

\paragraph{Thermally Grown Sample}

\begin{figure}[!hbt]
\includegraphics[width=0.43\textwidth]{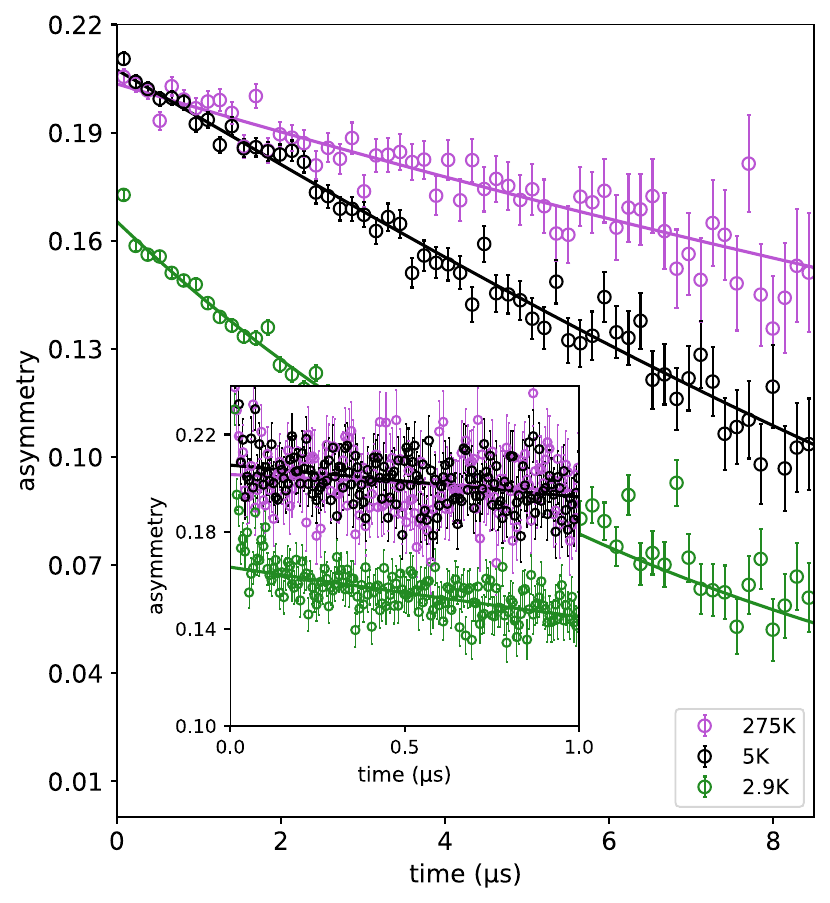}
\caption{Long time zero field $\mu$SR time spectra of thermal Ta$_2$O$_5$ (S1), for $E=2$~keV ($z=15$~nm) at $T=2.8$~K (green), $5$~K (black), and $275$~K (purple). inset: Short time zero field $\mu$SR time spectra of thermal Ta$_2$O$_5$ (S1). }\label{fig:Ta2O5_thermal_ZF_both}
\end{figure}

The ZF temperature dependence of the thermally grown films is very different from that of the sputtered ones. 
Fig.\ref{fig:Ta2O5_thermal_ZF_both} inset shows the short-time behavior for $E=2$~keV ($z=15$~nm), for various temperatures. Only at the lowest temperature, there is a hint of a ZF precession, though the damping is so strong that it cannot be reliably analyzed. The long time ZF time spectra in Fig.~\ref{fig:Ta2O5_thermal_ZF_both} show that between RT and $T=5$~K the initial asymmetry remains and there is only a gradual increase in the depolarization rate. Between $T=5$~K and $T=2.8$~K there is a substantial drop in the initial asymmetry. These features suggest that for the thermal films there is also a low-$T$ magnetic transition, although nothing is observed at high temperature, which fits the TF results presented in Fig.~\ref{fig:Ta_oxides_fM}.

\subsubsection{Longitudinal field results}

\begin{figure}[!htb]
\includegraphics[width=0.47\textwidth]{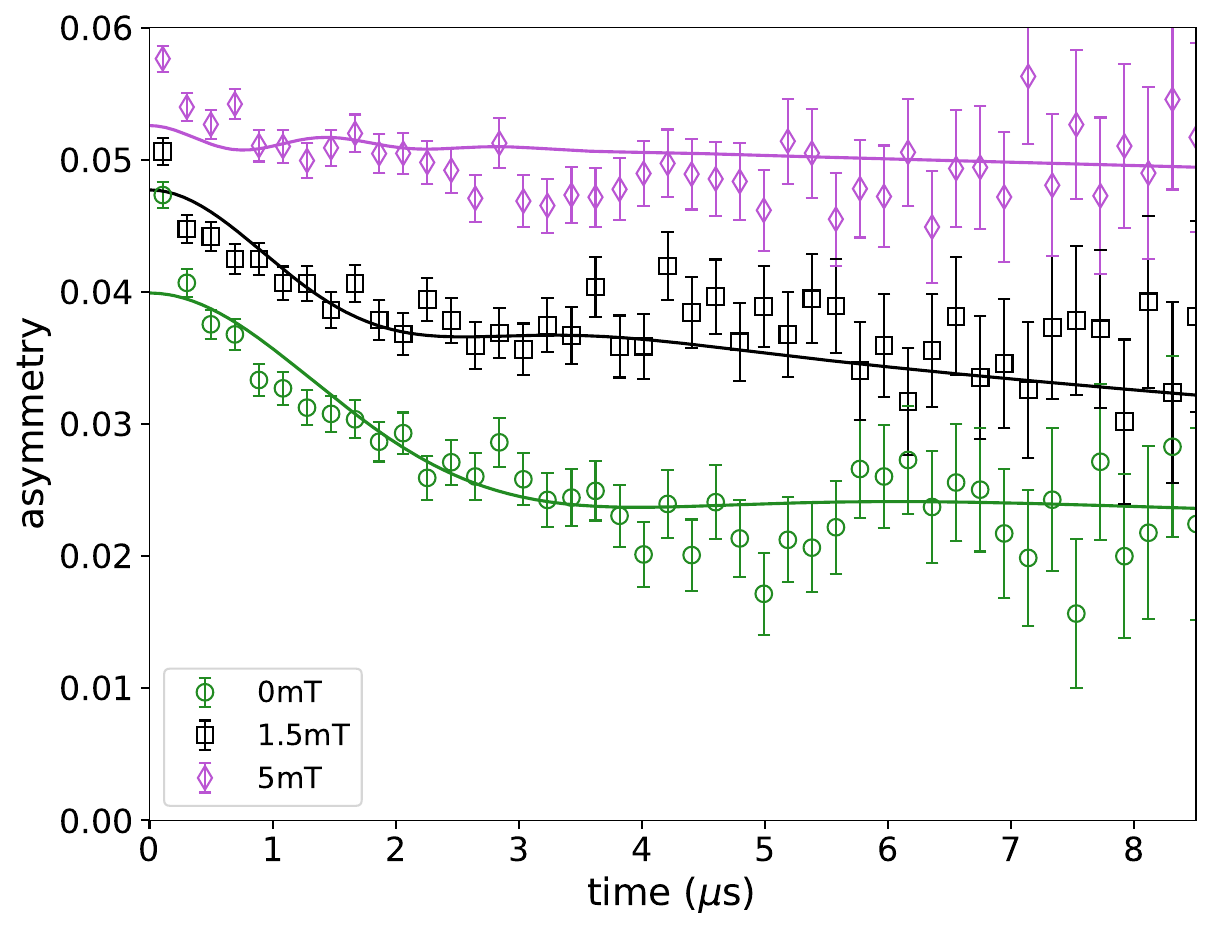}% Here is how to import EPS art
\caption{\label{fig:LF_TaOx}  $\mu$SR data in longitudinal magnetic field (LF) in sputtered Ta$_2$O$_5$, $E=2$~keV (15~nm) at base temperature $T=2.8$~K. Rate parameters for this global fit: $\nu$ = 0.29~$\mu\mathrm{s}^{-1}$, $\Delta$= 0.54~$\mu\mathrm{s}^{-1}$. }
\end{figure}

LF measurements were performed only for the sputtered films. Fig.~\ref{fig:LF_TaOx} shows the LF $B$-scan at $T=2.8$~K. The LF time spectrum is essentially flat at fields as low as $B=5$~mT, which suggests that in the predominant volume of the sample there are no magnetic fluctuations present.

\section{Discussion}

%new
The main observation from these experiments is that both amorphous niobium and tantalum pentoxides exhibit magnetism at low temperatures. However, the nature of this magnetism is fundamentally different between the two materials.

The microscopic origin of the observed magnetism in amorphous Nb$_2$O$_5$ and amorphous Ta$_2$O$_5$ is most likely due to oxygen vacancies presence. Oxygen vacancies ($V_{\rm O}$) are well established as triggers for magnetic behavior in various otherwise nonmagnetic materials, including HfO$_2$ and SrTiO$_3$ \cite{Venkatesan2004,Chaboy2010,Yang2010,ZHANG20121770}.

In amorphous Nb$_2$O$_5$, we observe a high level of magnetic disorder, where strong magnetic fluctuations prevent any form of ordering. In contrast, amorphous Ta$_2$O$_5$ shows signs of local static magnetism, suggesting some degree of order. This does not rule out the presence of fluctuations as they may simply lie outside the range of our measurements.

In the case of Nb$_2$O$_5$, the fluctuation rate is unexpectedly high ($\approx$~100 MHz), with a broad spectral range that likely overlaps the GHz regime. This overlap raises a critical concern: such fluctuations may interfere directly with the operational frequencies of superconducting qubits, contributing to decoherence. In other words, these fluctuations could be an unavoidable source of microwave loss in devices coupled to Nb$_2$O$_5$, and therefore, a limiting factor in superconducting device performance.

We can estimate the density of magnetic centers in Nb$_2$O$_5$ using relaxation rates from both longitudinal field (LF) and transverse field (TF) $\mu$SR measurements. From the fast fluctuation limit formula \cite{Yaouanc2011}:
\begin{equation}
    \lambda_{e}=\Delta^{2}/\nu,
\end{equation}
with $\lambda_{e}$ the depolarization rate caused by the presence of correlated electrons.
At base temperature (T = 2.8 K), our data gives:
\begin{equation}
    \Delta^{2}=0.2 {\mu s}^{-1}\cdot100 {\mu s}^{-1}=20 {\mu s}^{-2}.
\end{equation}
This value matches well with the $\Delta$ extracted from LF fits. The fluctuation amplitude is related to $\Delta$ via:

\begin{equation}
    \Delta^2=\gamma_{\mu}^2 \langle\delta B^2\rangle.
\end{equation}

\noindent Assuming a uniform distribution of magnetic defects and dipolar interaction:
\begin{equation}
    \Delta^2=\frac{4}{15} \hbar^2 {\gamma_{\mu}}^2 {\gamma_e}^2 S(S+1)\frac{1}{r^6}. 
\end{equation}

\noindent With assumption $S = 1/2$ for simplicity (magnetism could be defined by higher spin):
\begin{equation}\label{eq:2ndmom}
\begin{split}
    \Delta^2=\frac{4}{15} \hbar^2 {\gamma_{\mu}}^2 {\gamma_e}^2 S(S+1) \frac{1}{a^6}\Big|_{S=1/2} = \\
    =\frac{1}{5} \hbar^2 {\gamma_{\mu}}^2 {\gamma_e}^2 \frac{1}{a^6}.
\end{split}    
\end{equation}
Taking $a_0 = 2 \cdot 10^{-10}$ m as a characteristic distance between defects (roughly Nb–O bond length), we estimate an effective magnetic range of ~54 nm. Factoring in a volume fraction of 0.1, the defect density becomes $5 \cdot 10^{-9}$ per unit cell, corresponding to $3 \cdot 10^{15}$ defects per cm$^{3}$ in the bulk. Differences in frequency and temperature regimes likely explain why this value does not match Ref.\cite{Sendelbach2008} exactly.

Although static magnetism, especially from antiferromagnetically correlated defects, may appear less detrimental due to its zero net magnetization, it can still disturb superconductivity.  
Local magnetic fields remain nonzero and can affect nearby Cooper pairs through proximity, potentially leading to the formation of in-gap states, as previously observed in STM measurements on granular aluminum~\cite{Yang2020Sep} and tantalum~\cite{Arabi2024Dec} . 
In our case, the local mean field at the muon site reaches up to 150 mT, which is a substantial perturbation.

In amorphous Ta$_2$O$_5$, we did not observe high-frequency magnetic fluctuations, making it comparatively more favorable for superconducting environments. Furthermore, if the observed static magnetism is indeed due to antiferromagnetic correlations, this could offer an additional advantage in the case of finite-size clusters: antiferromagnetic systems often exhibit a finite excitation gap, suppressing low-energy spin dynamics. In such cases, magnetic excitations are expected to lie well above 10~GHz, beyond the operating frequencies of typical superconducting qubits. This absence of low-energy excitations could account for the relatively low magnetic noise observed in Ta$_2$O$_5$ and highlights the importance of excitation spectrum characteristics in material selection for superconducting quantum devices.

The $\mu$SR measurements of the thermal Ta$_2$O$_5$ films show essentially no magnetic volume fraction in the weak transverse field data for both samples, and a slight signature of a low temperature magnetic transition. In XRD measurements (see Appendix B) we clearly detected a certain amount of crystallinity in the thicker film S1. These diffraction peaks are absent for the thinner sample. However, since both S1 and S2 only differ in thickness, likely there is also some crystallinity in sample S2; the crystallites are just too small to detect. Consequently, this relatively ordered structure and a likely associated lower defect density is a possible explanation for the difference in magnetic behavior observed compared to the sputtered Ta$_2$O$_5$ film. Our data suggest that increasing crystallinity in Ta$_2$O$_5$ films is a valid option for suppressing magnetism and is thus beneficial in superconducting circuit applications. While this will be difficult to achieve for the surface oxide due to oxygen diffusion in Ta\cite{Chandrasekharan2005Dec}, it maybe possible in heterostructures for Josephson junctions.  

\section{Conclusions}

Our measurements reveal significant differences in the magnetic behavior of Nb$_2$O$_5$ and Ta$_2$O$_5$. While tantalum pentoxide (especially in its crystalline form) exhibits reduced magnetism and no observable GHz-range fluctuations, amorphous niobium pentoxide is magnetically disordered, with persistent, broadband fluctuations that likely extend into qubit operating frequencies. These GHz-scale fluctuations in Nb$_2$O$_5$ could be linked to various loss mechanisms, or could indicate local magnetic field suppression of the superconducting gap, acting as quasiparticle traps \cite{Graaf2020}. This highlights a key takeaway: magnetic disorder is strongly linked to increased microwave loss. We also observe that crystallinity plays a crucial role. Thermally grown (crystalline) Ta$_2$O$_5$ shows suppressed magnetism — though not completely zero — suggesting that improved control over crystallinity may be a practical path forward for optimizing material performance. Ideally, all oxides should be removed from the surfaces of superconducting devices to minimize loss. However, since this is not always practically feasible, our findings support recent studies \cite{Bal2024} and suggest that Ta-based systems may significantly outperform Nb-based ones in applications such as SQUIDs and flux-tunable qubits.

\section{Acknowledgments}

This work was supported by the U.S. Department of Energy, Office of Science, National Quantum Information Science Research Centers, Superconducting Quantum Materials and Systems Center (SQMS), under Contract No. 89243024CSC000002. Fermilab is operated by Fermi Forward Discovery Group, LLC under Contract No. 89243024CSC000002 with the U.S. Department of Energy, Office of Science, Office of High Energy Physics.
 This work made use of the S$\mu$S Facility of Paul Scherrer Institute. The authors thank for valuable discussions members of SQMS Center collaboration, in particular Ziwen Huang and Graham Pritchard. Valuable conversation authors had with Lara Faoro and Lev Ioffe and authors are thankful for their input.

T.R., R.D. and I.M.P. acknowledge funding from the Helmholtz Association, the European Union’s Horizon 2020 research and innovation programme under the Marie Skłodowska-Curie grant agreement number 847471, and the Federal Ministry of Education and Research via Project GeQCoS (FKZ: 13N15683). The authors are grateful to L. Radtke at KIT for technical support. Facilities use was supported by the KIT Nanostructure Service Laboratory (NSL) and the Karlsruhe Nano Micro Facility (KNMFi).

\nocite{*}

\bibliography{main}% Produces the bibliography via BibTeX.

\providecommand{\noopsort}[1]{}\providecommand{\singleletter}[1]{#1}%
\begin{thebibliography}{34}%
\makeatletter
\providecommand \@ifxundefined [1]{%
 \@ifx{#1\undefined}
}%
\providecommand \@ifnum [1]{%
 \ifnum #1\expandafter \@firstoftwo
 \else \expandafter \@secondoftwo
 \fi
}%
\providecommand \@ifx [1]{%
 \ifx #1\expandafter \@firstoftwo
 \else \expandafter \@secondoftwo
 \fi
}%
\providecommand \natexlab [1]{#1}%
\providecommand \enquote  [1]{``#1''}%
\providecommand \bibnamefont  [1]{#1}%
\providecommand \bibfnamefont [1]{#1}%
\providecommand \citenamefont [1]{#1}%
\providecommand \href@noop [0]{\@secondoftwo}%
\providecommand \href [0]{\begingroup \@sanitize@url \@href}%
\providecommand \@href[1]{\@@startlink{#1}\@@href}%
\providecommand \@@href[1]{\endgroup#1\@@endlink}%
\providecommand \@sanitize@url [0]{\catcode `\\12\catcode `\$12\catcode `\&12\catcode `\#12\catcode `\^12\catcode `\_12\catcode `\%12\relax}%
\providecommand \@@startlink[1]{}%
\providecommand \@@endlink[0]{}%
\providecommand \url  [0]{\begingroup\@sanitize@url \@url }%
\providecommand \@url [1]{\endgroup\@href {#1}{\urlprefix }}%
\providecommand \urlprefix  [0]{URL }%
\providecommand \Eprint [0]{\href }%
\providecommand \doibase [0]{https://doi.org/}%
\providecommand \selectlanguage [0]{\@gobble}%
\providecommand \bibinfo  [0]{\@secondoftwo}%
\providecommand \bibfield  [0]{\@secondoftwo}%
\providecommand \translation [1]{[#1]}%
\providecommand \BibitemOpen [0]{}%
\providecommand \bibitemStop [0]{}%
\providecommand \bibitemNoStop [0]{.\EOS\space}%
\providecommand \EOS [0]{\spacefactor3000\relax}%
\providecommand \BibitemShut  [1]{\csname bibitem#1\endcsname}%
\let\auto@bib@innerbib\@empty
%</preamble>
\bibitem [{\citenamefont {Krantz}\ \emph {et~al.}(2019)\citenamefont {Krantz}, \citenamefont {Kjaergaard}, \citenamefont {Yan}, \citenamefont {Orlando}, \citenamefont {Gustavsson},\ and\ \citenamefont {Oliver}}]{Krantz2019}%
  \BibitemOpen
  \bibfield  {author} {\bibinfo {author} {\bibfnamefont {P.}~\bibnamefont {Krantz}}, \bibinfo {author} {\bibfnamefont {M.}~\bibnamefont {Kjaergaard}}, \bibinfo {author} {\bibfnamefont {F.}~\bibnamefont {Yan}}, \bibinfo {author} {\bibfnamefont {T.~P.}\ \bibnamefont {Orlando}}, \bibinfo {author} {\bibfnamefont {S.}~\bibnamefont {Gustavsson}},\ and\ \bibinfo {author} {\bibfnamefont {W.~D.}\ \bibnamefont {Oliver}},\ }\bibfield  {title} {\bibinfo {title} {A quantum engineer's guide to superconducting qubits},\ }\href {https://doi.org/10.1063/1.5089550} {\bibfield  {journal} {\bibinfo  {journal} {Applied Physics Reviews}\ }\textbf {\bibinfo {volume} {6}},\ \bibinfo {pages} {021318} (\bibinfo {year} {2019})}\BibitemShut {NoStop}%
\bibitem [{\citenamefont {Romanenko}\ and\ \citenamefont {Schuster}(2017)}]{Romanenko_Schuster2017}%
  \BibitemOpen
  \bibfield  {author} {\bibinfo {author} {\bibfnamefont {A.}~\bibnamefont {Romanenko}}\ and\ \bibinfo {author} {\bibfnamefont {D.~I.}\ \bibnamefont {Schuster}},\ }\bibfield  {title} {\bibinfo {title} {Understanding quality factor degradation in superconducting niobium cavities at low microwave field amplitudes},\ }\href {https://doi.org/10.1103/PhysRevLett.119.264801} {\bibfield  {journal} {\bibinfo  {journal} {Phys. Rev. Lett.}\ }\textbf {\bibinfo {volume} {119}},\ \bibinfo {pages} {264801} (\bibinfo {year} {2017})}\BibitemShut {NoStop}%
\bibitem [{\citenamefont {Romanenko}\ \emph {et~al.}(2020)\citenamefont {Romanenko}, \citenamefont {Pilipenko}, \citenamefont {Zorzetti}, \citenamefont {Frolov}, \citenamefont {Awida}, \citenamefont {Belomestnykh}, \citenamefont {Posen},\ and\ \citenamefont {Grassellino}}]{Romanenko2020}%
  \BibitemOpen
  \bibfield  {author} {\bibinfo {author} {\bibfnamefont {A.}~\bibnamefont {Romanenko}}, \bibinfo {author} {\bibfnamefont {R.}~\bibnamefont {Pilipenko}}, \bibinfo {author} {\bibfnamefont {S.}~\bibnamefont {Zorzetti}}, \bibinfo {author} {\bibfnamefont {D.}~\bibnamefont {Frolov}}, \bibinfo {author} {\bibfnamefont {M.}~\bibnamefont {Awida}}, \bibinfo {author} {\bibfnamefont {S.}~\bibnamefont {Belomestnykh}}, \bibinfo {author} {\bibfnamefont {S.}~\bibnamefont {Posen}},\ and\ \bibinfo {author} {\bibfnamefont {A.}~\bibnamefont {Grassellino}},\ }\bibfield  {title} {\bibinfo {title} {Three-dimensional superconducting resonators at $t=20$ mk with photon lifetimes up to $\ensuremath{\tau}=2$ s},\ }\href {https://doi.org/10.1103/PhysRevApplied.13.034032} {\bibfield  {journal} {\bibinfo  {journal} {Phys. Rev. Appl.}\ }\textbf {\bibinfo {volume} {13}},\ \bibinfo {pages} {034032} (\bibinfo {year} {2020})}\BibitemShut {NoStop}%
\bibitem [{\citenamefont {Greener}\ \emph {et~al.}(1961)\citenamefont {Greener}, \citenamefont {Whitmore},\ and\ \citenamefont {Fine}}]{Greener1961}%
  \BibitemOpen
  \bibfield  {author} {\bibinfo {author} {\bibfnamefont {E.~H.}\ \bibnamefont {Greener}}, \bibinfo {author} {\bibfnamefont {D.~H.}\ \bibnamefont {Whitmore}},\ and\ \bibinfo {author} {\bibfnamefont {M.~E.}\ \bibnamefont {Fine}},\ }\bibfield  {title} {\bibinfo {title} {Electrical conductivity of near‐stoichiometric $\alpha$-nb$_2$o$_5$},\ }\href {https://doi.org/10.1063/1.1731627} {\bibfield  {journal} {\bibinfo  {journal} {The Journal of Chemical Physics}\ }\textbf {\bibinfo {volume} {34}},\ \bibinfo {pages} {1017} (\bibinfo {year} {1961})}\BibitemShut {NoStop}%
\bibitem [{\citenamefont {Streiff}\ \emph {et~al.}(1971)\citenamefont {Streiff}, \citenamefont {Poulton},\ and\ \citenamefont {Smeltzer}}]{Streiff1971}%
  \BibitemOpen
  \bibfield  {author} {\bibinfo {author} {\bibfnamefont {R.}~\bibnamefont {Streiff}}, \bibinfo {author} {\bibfnamefont {D.}~\bibnamefont {Poulton}},\ and\ \bibinfo {author} {\bibfnamefont {W.}~\bibnamefont {Smeltzer}},\ }\bibfield  {title} {\bibinfo {title} {Comment on the defect structure and the transport properties of nb$_2$o$_5$.},\ }\href {https://doi.org/https://doi.org/10.1007/BF00604738} {\bibfield  {journal} {\bibinfo  {journal} {Oxid Met}\ }\textbf {\bibinfo {volume} {3}},\ \bibinfo {pages} {33} (\bibinfo {year} {1971})}\BibitemShut {NoStop}%
\bibitem [{\citenamefont {Cava}\ \emph {et~al.}(1991)\citenamefont {Cava}, \citenamefont {Batlogg}, \citenamefont {Krajewski}, \citenamefont {Poulsen}, \citenamefont {Gammel}, \citenamefont {Peck},\ and\ \citenamefont {Rupp}}]{Cava1991}%
  \BibitemOpen
  \bibfield  {author} {\bibinfo {author} {\bibfnamefont {R.~J.}\ \bibnamefont {Cava}}, \bibinfo {author} {\bibfnamefont {B.}~\bibnamefont {Batlogg}}, \bibinfo {author} {\bibfnamefont {J.~J.}\ \bibnamefont {Krajewski}}, \bibinfo {author} {\bibfnamefont {H.~F.}\ \bibnamefont {Poulsen}}, \bibinfo {author} {\bibfnamefont {P.}~\bibnamefont {Gammel}}, \bibinfo {author} {\bibfnamefont {W.~F.}\ \bibnamefont {Peck}},\ and\ \bibinfo {author} {\bibfnamefont {L.~W.}\ \bibnamefont {Rupp}},\ }\bibfield  {title} {\bibinfo {title} {Electrical and magnetic properties of nb$_2$o$_{5-\delta}$ crystallographic shear structures},\ }\href {https://doi.org/10.1103/PhysRevB.44.6973} {\bibfield  {journal} {\bibinfo  {journal} {Phys. Rev. B}\ }\textbf {\bibinfo {volume} {44}},\ \bibinfo {pages} {6973} (\bibinfo {year} {1991})}\BibitemShut {NoStop}%
\bibitem [{\citenamefont {Proslier}\ \emph {et~al.}(2011)\citenamefont {Proslier}, \citenamefont {Kharitonov}, \citenamefont {Pellin}, \citenamefont {Zasadzinski},\ and\ \citenamefont {Ciovati}}]{Proslier2011}%
  \BibitemOpen
  \bibfield  {author} {\bibinfo {author} {\bibfnamefont {T.}~\bibnamefont {Proslier}}, \bibinfo {author} {\bibfnamefont {M.}~\bibnamefont {Kharitonov}}, \bibinfo {author} {\bibfnamefont {M.}~\bibnamefont {Pellin}}, \bibinfo {author} {\bibfnamefont {J.}~\bibnamefont {Zasadzinski}},\ and\ \bibinfo {author} {\bibnamefont {Ciovati}},\ }\bibfield  {title} {\bibinfo {title} {Evidence of surface paramagnetism in niobium and consequences for the superconducting cavity surface impedance},\ }\href {https://doi.org/10.1109/TASC.2011.2107491} {\bibfield  {journal} {\bibinfo  {journal} {IEEE Transactions on Applied Superconductivity}\ }\textbf {\bibinfo {volume} {21}},\ \bibinfo {pages} {2619} (\bibinfo {year} {2011})}\BibitemShut {NoStop}%
\bibitem [{\citenamefont {Wenskat}\ \emph {et~al.}(2022)\citenamefont {Wenskat}, \citenamefont {\ifmmode \check{C}\else \v{C}\fi{}i\ifmmode~\check{z}\else \v{z}\fi{}ek}, \citenamefont {Liedke},\ and\ \citenamefont {et~al}}]{Wenskat2022}%
  \BibitemOpen
  \bibfield  {author} {\bibinfo {author} {\bibfnamefont {M.}~\bibnamefont {Wenskat}}, \bibinfo {author} {\bibfnamefont {J.}~\bibnamefont {\ifmmode \check{C}\else \v{C}\fi{}i\ifmmode~\check{z}\else \v{z}\fi{}ek}}, \bibinfo {author} {\bibfnamefont {M.~O.}\ \bibnamefont {Liedke}},\ and\ \bibinfo {author} {\bibnamefont {et~al}},\ }\bibfield  {title} {\bibinfo {title} {Vacancy dynamics in niobium and its native oxides and their potential implications for quantum computing and superconducting accelerators},\ }\href {https://doi.org/10.1103/PhysRevB.106.094516} {\bibfield  {journal} {\bibinfo  {journal} {Phys. Rev. B}\ }\textbf {\bibinfo {volume} {106}},\ \bibinfo {pages} {094516} (\bibinfo {year} {2022})}\BibitemShut {NoStop}%
\bibitem [{\citenamefont {Murthy}\ \emph {et~al.}(2022)\citenamefont {Murthy}, \citenamefont {Masih~Das}, \citenamefont {Ribet}, \citenamefont {Kopas}, \citenamefont {Lee}, \citenamefont {Reagor}, \citenamefont {Zhou}, \citenamefont {Kramer}, \citenamefont {Hersam}, \citenamefont {Checchin}, \citenamefont {Grassellino}, \citenamefont {Reis}, \citenamefont {Dravid},\ and\ \citenamefont {Romanenko}}]{Murthy2022}%
  \BibitemOpen
  \bibfield  {author} {\bibinfo {author} {\bibfnamefont {A.~A.}\ \bibnamefont {Murthy}}, \bibinfo {author} {\bibfnamefont {P.}~\bibnamefont {Masih~Das}}, \bibinfo {author} {\bibfnamefont {S.~M.}\ \bibnamefont {Ribet}}, \bibinfo {author} {\bibfnamefont {C.}~\bibnamefont {Kopas}}, \bibinfo {author} {\bibfnamefont {J.}~\bibnamefont {Lee}}, \bibinfo {author} {\bibfnamefont {M.~J.}\ \bibnamefont {Reagor}}, \bibinfo {author} {\bibfnamefont {L.}~\bibnamefont {Zhou}}, \bibinfo {author} {\bibfnamefont {M.~J.}\ \bibnamefont {Kramer}}, \bibinfo {author} {\bibfnamefont {M.~C.}\ \bibnamefont {Hersam}}, \bibinfo {author} {\bibfnamefont {M.}~\bibnamefont {Checchin}}, \bibinfo {author} {\bibfnamefont {A.}~\bibnamefont {Grassellino}}, \bibinfo {author} {\bibfnamefont {R.~d.}\ \bibnamefont {Reis}}, \bibinfo {author} {\bibfnamefont {V.~P.}\ \bibnamefont {Dravid}},\ and\ \bibinfo {author} {\bibfnamefont {A.}~\bibnamefont {Romanenko}},\ }\bibfield  {title} {\bibinfo {title} {{Developing a Chemical and Structural Understanding of
  the Surface Oxide in a Niobium Superconducting Qubit}},\ }\href {https://doi.org/10.1021/acsnano.2c07913} {\bibfield  {journal} {\bibinfo  {journal} {ACS Nano}\ }\textbf {\bibinfo {volume} {16}},\ \bibinfo {pages} {17257} (\bibinfo {year} {2022})}\BibitemShut {NoStop}%
\bibitem [{\citenamefont {Bafia}\ \emph {et~al.}(2024)\citenamefont {Bafia}, \citenamefont {Murthy}, \citenamefont {Grassellino},\ and\ \citenamefont {Romanenko}}]{Bafia2024}%
  \BibitemOpen
  \bibfield  {author} {\bibinfo {author} {\bibfnamefont {D.}~\bibnamefont {Bafia}}, \bibinfo {author} {\bibfnamefont {A.}~\bibnamefont {Murthy}}, \bibinfo {author} {\bibfnamefont {A.}~\bibnamefont {Grassellino}},\ and\ \bibinfo {author} {\bibfnamefont {A.}~\bibnamefont {Romanenko}},\ }\bibfield  {title} {\bibinfo {title} {Oxygen vacancies in niobium pentoxide as a source of two-level system losses in superconducting niobium},\ }\href {https://doi.org/10.1103/PhysRevApplied.22.024035} {\bibfield  {journal} {\bibinfo  {journal} {Phys. Rev. Appl.}\ }\textbf {\bibinfo {volume} {22}},\ \bibinfo {pages} {024035} (\bibinfo {year} {2024})}\BibitemShut {NoStop}%
\bibitem [{\citenamefont {Wang}\ \emph {et~al.}(2022)\citenamefont {Wang}, \citenamefont {Li}, \citenamefont {Xu}, \citenamefont {Li}, \citenamefont {Wang}, \citenamefont {Yang}, \citenamefont {Mi}, \citenamefont {Liang}, \citenamefont {Su}, \citenamefont {Yang}, \citenamefont {Wang}, \citenamefont {Wang}, \citenamefont {Li}, \citenamefont {Chen}, \citenamefont {Li}, \citenamefont {Linghu}, \citenamefont {Han}, \citenamefont {Zhang}, \citenamefont {Feng}, \citenamefont {Song}, \citenamefont {Ma}, \citenamefont {Zhang}, \citenamefont {Wang}, \citenamefont {Zhao}, \citenamefont {Liu}, \citenamefont {Xue}, \citenamefont {Jin},\ and\ \citenamefont {Yu}}]{Wang2022}%
  \BibitemOpen
  \bibfield  {author} {\bibinfo {author} {\bibfnamefont {C.}~\bibnamefont {Wang}}, \bibinfo {author} {\bibfnamefont {X.}~\bibnamefont {Li}}, \bibinfo {author} {\bibfnamefont {H.}~\bibnamefont {Xu}}, \bibinfo {author} {\bibfnamefont {Z.}~\bibnamefont {Li}}, \bibinfo {author} {\bibfnamefont {J.}~\bibnamefont {Wang}}, \bibinfo {author} {\bibfnamefont {Z.}~\bibnamefont {Yang}}, \bibinfo {author} {\bibfnamefont {Z.}~\bibnamefont {Mi}}, \bibinfo {author} {\bibfnamefont {X.}~\bibnamefont {Liang}}, \bibinfo {author} {\bibfnamefont {T.}~\bibnamefont {Su}}, \bibinfo {author} {\bibfnamefont {C.}~\bibnamefont {Yang}}, \bibinfo {author} {\bibfnamefont {G.}~\bibnamefont {Wang}}, \bibinfo {author} {\bibfnamefont {W.}~\bibnamefont {Wang}}, \bibinfo {author} {\bibfnamefont {Y.}~\bibnamefont {Li}}, \bibinfo {author} {\bibfnamefont {M.}~\bibnamefont {Chen}}, \bibinfo {author} {\bibfnamefont {C.}~\bibnamefont {Li}}, \bibinfo {author} {\bibfnamefont {K.}~\bibnamefont {Linghu}}, \bibinfo {author} {\bibfnamefont {J.}~\bibnamefont
  {Han}}, \bibinfo {author} {\bibfnamefont {Y.}~\bibnamefont {Zhang}}, \bibinfo {author} {\bibfnamefont {Y.}~\bibnamefont {Feng}}, \bibinfo {author} {\bibfnamefont {Y.}~\bibnamefont {Song}}, \bibinfo {author} {\bibfnamefont {T.}~\bibnamefont {Ma}}, \bibinfo {author} {\bibfnamefont {J.}~\bibnamefont {Zhang}}, \bibinfo {author} {\bibfnamefont {R.}~\bibnamefont {Wang}}, \bibinfo {author} {\bibfnamefont {P.}~\bibnamefont {Zhao}}, \bibinfo {author} {\bibfnamefont {W.}~\bibnamefont {Liu}}, \bibinfo {author} {\bibfnamefont {G.}~\bibnamefont {Xue}}, \bibinfo {author} {\bibfnamefont {Y.}~\bibnamefont {Jin}},\ and\ \bibinfo {author} {\bibfnamefont {H.}~\bibnamefont {Yu}},\ }\bibfield  {title} {\bibinfo {title} {Towards practical quantum computers: transmon qubit with a lifetime approaching 0.5 milliseconds},\ }\href {https://doi.org/10.1038/s41534-021-00510-2} {\bibfield  {journal} {\bibinfo  {journal} {npj Quantum Information}\ }\textbf {\bibinfo {volume} {8}},\ \bibinfo {pages} {3} (\bibinfo {year}
  {2022})}\BibitemShut {NoStop}%
\bibitem [{\citenamefont {Place}\ \emph {et~al.}(2021)\citenamefont {Place}, \citenamefont {Rodgers}, \citenamefont {Mundada}, \citenamefont {Smitham}, \citenamefont {Fitzpatrick}, \citenamefont {Leng}, \citenamefont {Premkumar}, \citenamefont {Bryon}, \citenamefont {Vrajitoarea}, \citenamefont {Sussman}, \citenamefont {Cheng}, \citenamefont {Madhavan}, \citenamefont {Babla}, \citenamefont {Le}, \citenamefont {Gang}, \citenamefont {J{\"a}ck}, \citenamefont {Gyenis}, \citenamefont {Yao}, \citenamefont {Cava}, \citenamefont {de~Leon},\ and\ \citenamefont {Houck}}]{Place2021}%
  \BibitemOpen
  \bibfield  {author} {\bibinfo {author} {\bibfnamefont {A.~P.~M.}\ \bibnamefont {Place}}, \bibinfo {author} {\bibfnamefont {L.~V.~H.}\ \bibnamefont {Rodgers}}, \bibinfo {author} {\bibfnamefont {P.}~\bibnamefont {Mundada}}, \bibinfo {author} {\bibfnamefont {B.~M.}\ \bibnamefont {Smitham}}, \bibinfo {author} {\bibfnamefont {M.}~\bibnamefont {Fitzpatrick}}, \bibinfo {author} {\bibfnamefont {Z.}~\bibnamefont {Leng}}, \bibinfo {author} {\bibfnamefont {A.}~\bibnamefont {Premkumar}}, \bibinfo {author} {\bibfnamefont {J.}~\bibnamefont {Bryon}}, \bibinfo {author} {\bibfnamefont {A.}~\bibnamefont {Vrajitoarea}}, \bibinfo {author} {\bibfnamefont {S.}~\bibnamefont {Sussman}}, \bibinfo {author} {\bibfnamefont {G.}~\bibnamefont {Cheng}}, \bibinfo {author} {\bibfnamefont {T.}~\bibnamefont {Madhavan}}, \bibinfo {author} {\bibfnamefont {H.~K.}\ \bibnamefont {Babla}}, \bibinfo {author} {\bibfnamefont {X.~H.}\ \bibnamefont {Le}}, \bibinfo {author} {\bibfnamefont {Y.}~\bibnamefont {Gang}}, \bibinfo {author} {\bibfnamefont
  {B.}~\bibnamefont {J{\"a}ck}}, \bibinfo {author} {\bibfnamefont {A.}~\bibnamefont {Gyenis}}, \bibinfo {author} {\bibfnamefont {N.}~\bibnamefont {Yao}}, \bibinfo {author} {\bibfnamefont {R.~J.}\ \bibnamefont {Cava}}, \bibinfo {author} {\bibfnamefont {N.~P.}\ \bibnamefont {de~Leon}},\ and\ \bibinfo {author} {\bibfnamefont {A.~A.}\ \bibnamefont {Houck}},\ }\bibfield  {title} {\bibinfo {title} {New material platform for superconducting transmon qubits with coherence times exceeding 0.3 milliseconds},\ }\href {https://doi.org/10.1038/s41467-021-22030-5} {\bibfield  {journal} {\bibinfo  {journal} {Nature Communications}\ }\textbf {\bibinfo {volume} {12}},\ \bibinfo {pages} {1779} (\bibinfo {year} {2021})}\BibitemShut {NoStop}%
\bibitem [{\citenamefont {Bland}\ \emph {et~al.}(2025)\citenamefont {Bland}, \citenamefont {Bahrami}, \citenamefont {Martinez}, \citenamefont {Prestegaard}, \citenamefont {Smitham}, \citenamefont {Joshi}, \citenamefont {Hedrick}, \citenamefont {Kumar}, \citenamefont {Yang}, \citenamefont {Pakpour-Tabrizi} \emph {et~al.}}]{bland2025millisecond}%
  \BibitemOpen
  \bibfield  {author} {\bibinfo {author} {\bibfnamefont {M.~P.}\ \bibnamefont {Bland}}, \bibinfo {author} {\bibfnamefont {F.}~\bibnamefont {Bahrami}}, \bibinfo {author} {\bibfnamefont {J.~G.}\ \bibnamefont {Martinez}}, \bibinfo {author} {\bibfnamefont {P.~H.}\ \bibnamefont {Prestegaard}}, \bibinfo {author} {\bibfnamefont {B.~M.}\ \bibnamefont {Smitham}}, \bibinfo {author} {\bibfnamefont {A.}~\bibnamefont {Joshi}}, \bibinfo {author} {\bibfnamefont {E.}~\bibnamefont {Hedrick}}, \bibinfo {author} {\bibfnamefont {S.}~\bibnamefont {Kumar}}, \bibinfo {author} {\bibfnamefont {A.}~\bibnamefont {Yang}}, \bibinfo {author} {\bibfnamefont {A.~C.}\ \bibnamefont {Pakpour-Tabrizi}}, \emph {et~al.},\ }\bibfield  {title} {\bibinfo {title} {Millisecond lifetimes and coherence times in 2d transmon qubits},\ }\href@noop {} {\bibfield  {journal} {\bibinfo  {journal} {Nature}\ ,\ \bibinfo {pages} {1}} (\bibinfo {year} {2025})}\BibitemShut {NoStop}%
\bibitem [{\citenamefont {Bal}\ \emph {et~al.}(2024)\citenamefont {Bal}, \citenamefont {Murthy}, \citenamefont {Zhu}, \citenamefont {Crisa}, \citenamefont {You}, \citenamefont {Huang}, \citenamefont {Roy}, \citenamefont {Lee}, \citenamefont {Zanten}, \citenamefont {Pilipenko}, \citenamefont {Nekrashevich}, \citenamefont {Lunin}, \citenamefont {Bafia}, \citenamefont {Krasnikova}, \citenamefont {Kopas}, \citenamefont {Lachman}, \citenamefont {Miller}, \citenamefont {Mutus}, \citenamefont {Reagor}, \citenamefont {Cansizoglu}, \citenamefont {Marshall}, \citenamefont {Pappas}, \citenamefont {Vu}, \citenamefont {Yadavalli}, \citenamefont {Oh}, \citenamefont {Zhou}, \citenamefont {Kramer}, \citenamefont {Lecocq}, \citenamefont {Goronzy}, \citenamefont {Torres-Castanedo}, \citenamefont {Pritchard}, \citenamefont {Dravid}, \citenamefont {Rondinelli}, \citenamefont {Bedzyk}, \citenamefont {Hersam}, \citenamefont {Zasadzinski}, \citenamefont {Koch}, \citenamefont {Sauls}, \citenamefont {Romanenko},\ and\ \citenamefont
  {Grassellino}}]{Bal2024}%
  \BibitemOpen
  \bibfield  {author} {\bibinfo {author} {\bibfnamefont {M.}~\bibnamefont {Bal}}, \bibinfo {author} {\bibfnamefont {A.~A.}\ \bibnamefont {Murthy}}, \bibinfo {author} {\bibfnamefont {S.}~\bibnamefont {Zhu}}, \bibinfo {author} {\bibfnamefont {F.}~\bibnamefont {Crisa}}, \bibinfo {author} {\bibfnamefont {X.}~\bibnamefont {You}}, \bibinfo {author} {\bibfnamefont {Z.}~\bibnamefont {Huang}}, \bibinfo {author} {\bibfnamefont {T.}~\bibnamefont {Roy}}, \bibinfo {author} {\bibfnamefont {J.}~\bibnamefont {Lee}}, \bibinfo {author} {\bibfnamefont {D.~v.}\ \bibnamefont {Zanten}}, \bibinfo {author} {\bibfnamefont {R.}~\bibnamefont {Pilipenko}}, \bibinfo {author} {\bibfnamefont {I.}~\bibnamefont {Nekrashevich}}, \bibinfo {author} {\bibfnamefont {A.}~\bibnamefont {Lunin}}, \bibinfo {author} {\bibfnamefont {D.}~\bibnamefont {Bafia}}, \bibinfo {author} {\bibfnamefont {Y.}~\bibnamefont {Krasnikova}}, \bibinfo {author} {\bibfnamefont {C.~J.}\ \bibnamefont {Kopas}}, \bibinfo {author} {\bibfnamefont {E.~O.}\ \bibnamefont {Lachman}},
  \bibinfo {author} {\bibfnamefont {D.}~\bibnamefont {Miller}}, \bibinfo {author} {\bibfnamefont {J.~Y.}\ \bibnamefont {Mutus}}, \bibinfo {author} {\bibfnamefont {M.~J.}\ \bibnamefont {Reagor}}, \bibinfo {author} {\bibfnamefont {H.}~\bibnamefont {Cansizoglu}}, \bibinfo {author} {\bibfnamefont {J.}~\bibnamefont {Marshall}}, \bibinfo {author} {\bibfnamefont {D.~P.}\ \bibnamefont {Pappas}}, \bibinfo {author} {\bibfnamefont {K.}~\bibnamefont {Vu}}, \bibinfo {author} {\bibfnamefont {K.}~\bibnamefont {Yadavalli}}, \bibinfo {author} {\bibfnamefont {J.-S.}\ \bibnamefont {Oh}}, \bibinfo {author} {\bibfnamefont {L.}~\bibnamefont {Zhou}}, \bibinfo {author} {\bibfnamefont {M.~J.}\ \bibnamefont {Kramer}}, \bibinfo {author} {\bibfnamefont {F.}~\bibnamefont {Lecocq}}, \bibinfo {author} {\bibfnamefont {D.~P.}\ \bibnamefont {Goronzy}}, \bibinfo {author} {\bibfnamefont {C.~G.}\ \bibnamefont {Torres-Castanedo}}, \bibinfo {author} {\bibfnamefont {P.~G.}\ \bibnamefont {Pritchard}}, \bibinfo {author} {\bibfnamefont {V.~P.}\
  \bibnamefont {Dravid}}, \bibinfo {author} {\bibfnamefont {J.~M.}\ \bibnamefont {Rondinelli}}, \bibinfo {author} {\bibfnamefont {M.~J.}\ \bibnamefont {Bedzyk}}, \bibinfo {author} {\bibfnamefont {M.~C.}\ \bibnamefont {Hersam}}, \bibinfo {author} {\bibfnamefont {J.}~\bibnamefont {Zasadzinski}}, \bibinfo {author} {\bibfnamefont {J.}~\bibnamefont {Koch}}, \bibinfo {author} {\bibfnamefont {J.~A.}\ \bibnamefont {Sauls}}, \bibinfo {author} {\bibfnamefont {A.}~\bibnamefont {Romanenko}},\ and\ \bibinfo {author} {\bibfnamefont {A.}~\bibnamefont {Grassellino}},\ }\bibfield  {title} {\bibinfo {title} {Systematic improvements in transmon qubit coherence enabled by niobium surface encapsulation},\ }\href {https://doi.org/10.1038/s41534-024-00840-x} {\bibfield  {journal} {\bibinfo  {journal} {npj Quantum Information}\ }\textbf {\bibinfo {volume} {10}},\ \bibinfo {pages} {43} (\bibinfo {year} {2024})}\BibitemShut {NoStop}%
\bibitem [{\citenamefont {Pritchard}\ and\ \citenamefont {Rondinelli}(2025)}]{pritchard2025suppressed}%
  \BibitemOpen
  \bibfield  {author} {\bibinfo {author} {\bibfnamefont {P.~G.}\ \bibnamefont {Pritchard}}\ and\ \bibinfo {author} {\bibfnamefont {J.~M.}\ \bibnamefont {Rondinelli}},\ }\bibfield  {title} {\bibinfo {title} {Suppressed paramagnetism in amorphous ta 2 o 5- x oxides and its link to superconducting-qubit performance},\ }\href@noop {} {\bibfield  {journal} {\bibinfo  {journal} {Physical Review Applied}\ }\textbf {\bibinfo {volume} {23}},\ \bibinfo {pages} {064062} (\bibinfo {year} {2025})}\BibitemShut {NoStop}%
\bibitem [{\citenamefont {Blundell}\ \emph {et~al.}(2021)\citenamefont {Blundell}, \citenamefont {De~Renzi}, \citenamefont {Lancaster},\ and\ \citenamefont {Pratt}}]{Blundell2021}%
  \BibitemOpen
  \bibfield  {author} {\bibinfo {author} {\bibfnamefont {S.~J.}\ \bibnamefont {Blundell}}, \bibinfo {author} {\bibfnamefont {R.}~\bibnamefont {De~Renzi}}, \bibinfo {author} {\bibfnamefont {T.}~\bibnamefont {Lancaster}},\ and\ \bibinfo {author} {\bibfnamefont {F.~L.}\ \bibnamefont {Pratt}},\ }\bibfield  {title} {\bibinfo {title} {{Low energy µSR}},\ }in\ \href {https://doi.org/10.1093/oso/9780198858959.003.0018} {\emph {\bibinfo {booktitle} {{Muon Spectroscopy: An Introduction}}}}\ (\bibinfo  {publisher} {Oxford University Press},\ \bibinfo {year} {2021})\BibitemShut {NoStop}%
\bibitem [{\citenamefont {Morenzoni}\ \emph {et~al.}()\citenamefont {Morenzoni}, \citenamefont {Prokscha}, \citenamefont {Saadaoui}, \citenamefont {Salman}, \citenamefont {Suter}, \citenamefont {Wojek}, \citenamefont {Baglo}, \citenamefont {Božović}, \citenamefont {Hossain}, \citenamefont {Kiefl}, \citenamefont {Logvenov},\ and\ \citenamefont {Ofer}}]{Morenzoni2014}%
  \BibitemOpen
  \bibfield  {author} {\bibinfo {author} {\bibfnamefont {E.}~\bibnamefont {Morenzoni}}, \bibinfo {author} {\bibfnamefont {T.}~\bibnamefont {Prokscha}}, \bibinfo {author} {\bibfnamefont {H.}~\bibnamefont {Saadaoui}}, \bibinfo {author} {\bibfnamefont {Z.}~\bibnamefont {Salman}}, \bibinfo {author} {\bibfnamefont {A.}~\bibnamefont {Suter}}, \bibinfo {author} {\bibfnamefont {B.~M.}\ \bibnamefont {Wojek}}, \bibinfo {author} {\bibfnamefont {J.}~\bibnamefont {Baglo}}, \bibinfo {author} {\bibfnamefont {I.}~\bibnamefont {Božović}}, \bibinfo {author} {\bibfnamefont {M.}~\bibnamefont {Hossain}}, \bibinfo {author} {\bibfnamefont {R.~F.}\ \bibnamefont {Kiefl}}, \bibinfo {author} {\bibfnamefont {G.}~\bibnamefont {Logvenov}},\ and\ \bibinfo {author} {\bibfnamefont {O.}~\bibnamefont {Ofer}},\ }\href {https://doi.org/10.7566/JPSCP.2.010201} {\emph {\bibinfo {title} {Proceedings of the International Symposium on Science Explored by Ultra Slow Muon (USM2013)}}}\BibitemShut {NoStop}%
\bibitem [{\citenamefont {Prokscha}\ \emph {et~al.}(2008)\citenamefont {Prokscha}, \citenamefont {Morenzoni}, \citenamefont {Deiters}, \citenamefont {Foroughi}, \citenamefont {George}, \citenamefont {Kobler}, \citenamefont {Suter},\ and\ \citenamefont {Vrankovic}}]{Prokscha_2008}%
  \BibitemOpen
  \bibfield  {author} {\bibinfo {author} {\bibfnamefont {T.}~\bibnamefont {Prokscha}}, \bibinfo {author} {\bibfnamefont {E.}~\bibnamefont {Morenzoni}}, \bibinfo {author} {\bibfnamefont {K.}~\bibnamefont {Deiters}}, \bibinfo {author} {\bibfnamefont {F.}~\bibnamefont {Foroughi}}, \bibinfo {author} {\bibfnamefont {D.}~\bibnamefont {George}}, \bibinfo {author} {\bibfnamefont {R.}~\bibnamefont {Kobler}}, \bibinfo {author} {\bibfnamefont {A.}~\bibnamefont {Suter}},\ and\ \bibinfo {author} {\bibfnamefont {V.}~\bibnamefont {Vrankovic}},\ }\bibfield  {title} {\bibinfo {title} {The new {muE4} beam at {PSI}: {A} hybrid-type large acceptance channel for the generation of a high intensity surface-muon beam},\ }\href {https://doi.org/10.1016/j.nima.2008.07.081} {\bibfield  {journal} {\bibinfo  {journal} {Nuclear Instruments and Methods in Physics Research Section A: Accelerators, Spectrometers, Detectors and Associated Equipment}\ }\textbf {\bibinfo {volume} {595}},\ \bibinfo {pages} {317} (\bibinfo {year}
  {2008})}\BibitemShut {NoStop}%
\bibitem [{\citenamefont {W.}(1991)}]{Eckstein1991}%
  \BibitemOpen
  \bibfield  {author} {\bibinfo {author} {\bibfnamefont {E.}~\bibnamefont {W.}},\ }\href@noop {} {\emph {\bibinfo {title} {Computer Simulation of Ion-Solid Interactions}}}\ (\bibinfo {year} {1991})\BibitemShut {NoStop}%
\bibitem [{\citenamefont {Yaouanc}\ and\ \citenamefont {Dalmas~de Reotier}(2011)}]{Yaouanc2011}%
  \BibitemOpen
  \bibfield  {author} {\bibinfo {author} {\bibfnamefont {A.}~\bibnamefont {Yaouanc}}\ and\ \bibinfo {author} {\bibfnamefont {P.}~\bibnamefont {Dalmas~de Reotier}},\ }\href@noop {} {\emph {\bibinfo {title} {Muon spin rotation, relaxation, and resonance: Applications to condensed matter}}},\ Vol.\ \bibinfo {volume} {147}\ (\bibinfo {year} {2011})\BibitemShut {NoStop}%
\bibitem [{\citenamefont {Chandrasekharan}\ \emph {et~al.}(2005)\citenamefont {Chandrasekharan}, \citenamefont {Park}, \citenamefont {Masel},\ and\ \citenamefont {Shannon}}]{Chandrasekharan2005Dec}%
  \BibitemOpen
  \bibfield  {author} {\bibinfo {author} {\bibfnamefont {R.}~\bibnamefont {Chandrasekharan}}, \bibinfo {author} {\bibfnamefont {I.}~\bibnamefont {Park}}, \bibinfo {author} {\bibfnamefont {R.~I.}\ \bibnamefont {Masel}},\ and\ \bibinfo {author} {\bibfnamefont {M.~A.}\ \bibnamefont {Shannon}},\ }\bibfield  {title} {\bibinfo {title} {{Thermal oxidation of tantalum films at various oxidation states from 300 to 700{\ifmmode\mbox{\textdegree}\else\textdegree\fi}C}},\ }\href {https://doi.org/10.1063/1.2139834} {\bibfield  {journal} {\bibinfo  {journal} {J. Appl. Phys.}\ }\textbf {\bibinfo {volume} {98}},\ \bibinfo {pages} {114908} (\bibinfo {year} {2005})}\BibitemShut {NoStop}%
\bibitem [{\citenamefont {Suter}\ and\ \citenamefont {Wojek}(2012)}]{Suter2012musrfit}%
  \BibitemOpen
  \bibfield  {author} {\bibinfo {author} {\bibfnamefont {A.}~\bibnamefont {Suter}}\ and\ \bibinfo {author} {\bibfnamefont {B.}~\bibnamefont {Wojek}},\ }\bibfield  {title} {\bibinfo {title} {Musrfit: a free platform-independent framework for $\mu$sr data analysis},\ }\href {https://doi.org/10.1016/j.phpro.2012.04.042} {\bibfield  {journal} {\bibinfo  {journal} {Physics Procedia}\ }\textbf {\bibinfo {volume} {30}},\ \bibinfo {pages} {69} (\bibinfo {year} {2012})}\BibitemShut {NoStop}%
\bibitem [{\citenamefont {Suter}\ \emph {et~al.}(2023)\citenamefont {Suter}, \citenamefont {Martins}, \citenamefont {Ni}, \citenamefont {Prokscha},\ and\ \citenamefont {Salman}}]{Suter2023}%
  \BibitemOpen
  \bibfield  {author} {\bibinfo {author} {\bibfnamefont {A.}~\bibnamefont {Suter}}, \bibinfo {author} {\bibfnamefont {M.~M.}\ \bibnamefont {Martins}}, \bibinfo {author} {\bibfnamefont {X.}~\bibnamefont {Ni}}, \bibinfo {author} {\bibfnamefont {T.}~\bibnamefont {Prokscha}},\ and\ \bibinfo {author} {\bibfnamefont {Z.}~\bibnamefont {Salman}},\ }\bibfield  {title} {\bibinfo {title} {Low energy measurements in low-energy µsr},\ }\href {https://doi.org/10.1088/1742-6596/2462/1/012011} {\bibfield  {journal} {\bibinfo  {journal} {Journal of Physics: Conference Series}\ }\textbf {\bibinfo {volume} {2462}},\ \bibinfo {pages} {012011} (\bibinfo {year} {2023})}\BibitemShut {NoStop}%
\bibitem [{\citenamefont {Venkatesan}\ \emph {et~al.}(2004)\citenamefont {Venkatesan}, \citenamefont {Fitzgerald},\ and\ \citenamefont {Coey}}]{Venkatesan2004}%
  \BibitemOpen
  \bibfield  {author} {\bibinfo {author} {\bibfnamefont {M.}~\bibnamefont {Venkatesan}}, \bibinfo {author} {\bibfnamefont {C.~B.}\ \bibnamefont {Fitzgerald}},\ and\ \bibinfo {author} {\bibfnamefont {J.~M.~D.}\ \bibnamefont {Coey}},\ }\bibfield  {title} {\bibinfo {title} {Unexpected magnetism in a dielectric oxide},\ }\href {https://doi.org/10.1038/430630a} {\bibfield  {journal} {\bibinfo  {journal} {Nature}\ }\textbf {\bibinfo {volume} {430}},\ \bibinfo {pages} {630} (\bibinfo {year} {2004})}\BibitemShut {NoStop}%
\bibitem [{\citenamefont {Chaboy}\ \emph {et~al.}(2010)\citenamefont {Chaboy}, \citenamefont {Boada}, \citenamefont {Piquer}, \citenamefont {Laguna-Marco}, \citenamefont {Garc\'{\i}a-Hern\'andez}, \citenamefont {Carmona}, \citenamefont {Llopis}, \citenamefont {Ru\'{\i}z-Gonz\'alez}, \citenamefont {Gonz\'alez-Calbet}, \citenamefont {Fern\'andez},\ and\ \citenamefont {Garc\'{\i}a}}]{Chaboy2010}%
  \BibitemOpen
  \bibfield  {author} {\bibinfo {author} {\bibfnamefont {J.}~\bibnamefont {Chaboy}}, \bibinfo {author} {\bibfnamefont {R.}~\bibnamefont {Boada}}, \bibinfo {author} {\bibfnamefont {C.}~\bibnamefont {Piquer}}, \bibinfo {author} {\bibfnamefont {M.~A.}\ \bibnamefont {Laguna-Marco}}, \bibinfo {author} {\bibfnamefont {M.}~\bibnamefont {Garc\'{\i}a-Hern\'andez}}, \bibinfo {author} {\bibfnamefont {N.}~\bibnamefont {Carmona}}, \bibinfo {author} {\bibfnamefont {J.}~\bibnamefont {Llopis}}, \bibinfo {author} {\bibfnamefont {M.~L.}\ \bibnamefont {Ru\'{\i}z-Gonz\'alez}}, \bibinfo {author} {\bibfnamefont {J.}~\bibnamefont {Gonz\'alez-Calbet}}, \bibinfo {author} {\bibfnamefont {J.~F.}\ \bibnamefont {Fern\'andez}},\ and\ \bibinfo {author} {\bibfnamefont {M.~A.}\ \bibnamefont {Garc\'{\i}a}},\ }\bibfield  {title} {\bibinfo {title} {Evidence of intrinsic magnetism in capped zno nanoparticles},\ }\href {https://doi.org/10.1103/PhysRevB.82.064411} {\bibfield  {journal} {\bibinfo  {journal} {Phys. Rev. B}\ }\textbf {\bibinfo
  {volume} {82}},\ \bibinfo {pages} {064411} (\bibinfo {year} {2010})}\BibitemShut {NoStop}%
\bibitem [{\citenamefont {Yang}\ \emph {et~al.}(2010)\citenamefont {Yang}, \citenamefont {Dai}, \citenamefont {Huang},\ and\ \citenamefont {Feng}}]{Yang2010}%
  \BibitemOpen
  \bibfield  {author} {\bibinfo {author} {\bibfnamefont {K.}~\bibnamefont {Yang}}, \bibinfo {author} {\bibfnamefont {Y.}~\bibnamefont {Dai}}, \bibinfo {author} {\bibfnamefont {B.}~\bibnamefont {Huang}},\ and\ \bibinfo {author} {\bibfnamefont {Y.~P.}\ \bibnamefont {Feng}},\ }\bibfield  {title} {\bibinfo {title} {Density-functional characterization of antiferromagnetism in oxygen-deficient anatase and rutile ${\text{tio}}_{2}$},\ }\href {https://doi.org/10.1103/PhysRevB.81.033202} {\bibfield  {journal} {\bibinfo  {journal} {Phys. Rev. B}\ }\textbf {\bibinfo {volume} {81}},\ \bibinfo {pages} {033202} (\bibinfo {year} {2010})}\BibitemShut {NoStop}%
\bibitem [{\citenamefont {Zhang}\ \emph {et~al.}(2012)\citenamefont {Zhang}, \citenamefont {Hu}, \citenamefont {Cao}, \citenamefont {Sun},\ and\ \citenamefont {Qin}}]{ZHANG20121770}%
  \BibitemOpen
  \bibfield  {author} {\bibinfo {author} {\bibfnamefont {Y.}~\bibnamefont {Zhang}}, \bibinfo {author} {\bibfnamefont {J.}~\bibnamefont {Hu}}, \bibinfo {author} {\bibfnamefont {E.}~\bibnamefont {Cao}}, \bibinfo {author} {\bibfnamefont {L.}~\bibnamefont {Sun}},\ and\ \bibinfo {author} {\bibfnamefont {H.}~\bibnamefont {Qin}},\ }\bibfield  {title} {\bibinfo {title} {Vacancy induced magnetism in srtio3},\ }\href {https://doi.org/https://doi.org/10.1016/j.jmmm.2011.12.036} {\bibfield  {journal} {\bibinfo  {journal} {Journal of Magnetism and Magnetic Materials}\ }\textbf {\bibinfo {volume} {324}},\ \bibinfo {pages} {1770} (\bibinfo {year} {2012})}\BibitemShut {NoStop}%
\bibitem [{\citenamefont {Sendelbach}\ \emph {et~al.}(2008)\citenamefont {Sendelbach}, \citenamefont {Hover}, \citenamefont {Kittel}, \citenamefont {M{\"u}ck}, \citenamefont {Martinis},\ and\ \citenamefont {McDermott}}]{Sendelbach2008}%
  \BibitemOpen
  \bibfield  {author} {\bibinfo {author} {\bibfnamefont {S.}~\bibnamefont {Sendelbach}}, \bibinfo {author} {\bibfnamefont {D.}~\bibnamefont {Hover}}, \bibinfo {author} {\bibfnamefont {A.}~\bibnamefont {Kittel}}, \bibinfo {author} {\bibfnamefont {M.}~\bibnamefont {M{\"u}ck}}, \bibinfo {author} {\bibfnamefont {J.~M.}\ \bibnamefont {Martinis}},\ and\ \bibinfo {author} {\bibfnamefont {R.}~\bibnamefont {McDermott}},\ }\bibfield  {title} {\bibinfo {title} {Magnetism in squids at millikelvin temperatures},\ }\href {https://doi.org/10.1103/PhysRevLett.100.227006} {\bibfield  {journal} {\bibinfo  {journal} {Phys. Rev. Lett.}\ }\textbf {\bibinfo {volume} {100}},\ \bibinfo {pages} {227006} (\bibinfo {year} {2008})}\BibitemShut {NoStop}%
\bibitem [{\citenamefont {Yang}\ \emph {et~al.}(2020)\citenamefont {Yang}, \citenamefont {Gozlinski}, \citenamefont {Storbeck}, \citenamefont {Gr{\ifmmode\ddot{u}\else\"{u}\fi}nhaupt}, \citenamefont {Pop},\ and\ \citenamefont {Wulfhekel}}]{Yang2020Sep}%
  \BibitemOpen
  \bibfield  {author} {\bibinfo {author} {\bibfnamefont {F.}~\bibnamefont {Yang}}, \bibinfo {author} {\bibfnamefont {T.}~\bibnamefont {Gozlinski}}, \bibinfo {author} {\bibfnamefont {T.}~\bibnamefont {Storbeck}}, \bibinfo {author} {\bibfnamefont {L.}~\bibnamefont {Gr{\ifmmode\ddot{u}\else\"{u}\fi}nhaupt}}, \bibinfo {author} {\bibfnamefont {I.~M.}\ \bibnamefont {Pop}},\ and\ \bibinfo {author} {\bibfnamefont {W.}~\bibnamefont {Wulfhekel}},\ }\bibfield  {title} {\bibinfo {title} {{Microscopic charging and in-gap states in superconducting granular aluminum}},\ }\href {https://doi.org/10.1103/PhysRevB.102.104502} {\bibfield  {journal} {\bibinfo  {journal} {Phys. Rev. B}\ }\textbf {\bibinfo {volume} {102}},\ \bibinfo {pages} {104502} (\bibinfo {year} {2020})}\BibitemShut {NoStop}%
\bibitem [{\citenamefont {Arabi}\ \emph {et~al.}(2024)\citenamefont {Arabi}, \citenamefont {Li}, \citenamefont {Dhundhwal}, \citenamefont {Fuchs}, \citenamefont {Reisinger}, \citenamefont {Pop},\ and\ \citenamefont {Wulfhekel}}]{Arabi2024Dec}%
  \BibitemOpen
  \bibfield  {author} {\bibinfo {author} {\bibfnamefont {S.}~\bibnamefont {Arabi}}, \bibinfo {author} {\bibfnamefont {Q.}~\bibnamefont {Li}}, \bibinfo {author} {\bibfnamefont {R.}~\bibnamefont {Dhundhwal}}, \bibinfo {author} {\bibfnamefont {D.}~\bibnamefont {Fuchs}}, \bibinfo {author} {\bibfnamefont {T.}~\bibnamefont {Reisinger}}, \bibinfo {author} {\bibfnamefont {I.~M.}\ \bibnamefont {Pop}},\ and\ \bibinfo {author} {\bibfnamefont {W.}~\bibnamefont {Wulfhekel}},\ }\bibfield  {title} {\bibinfo {title} {{Magnetic bound states embedded in tantalum superconducting thin films}},\ }\bibfield  {journal} {\bibinfo  {journal} {arXiv}\ }\href {https://doi.org/10.1063/5.0251996} {10.1063/5.0251996} (\bibinfo {year} {2024}),\ \Eprint {https://arxiv.org/abs/2412.15903} {2412.15903} \BibitemShut {NoStop}%
\bibitem [{\citenamefont {de~Graaf}\ \emph {et~al.}(2020)\citenamefont {de~Graaf}, \citenamefont {Faoro}, \citenamefont {Ioffe}, \citenamefont {Mahashabde}, \citenamefont {Burnett}, \citenamefont {Lindström}, \citenamefont {Kubatkin}, \citenamefont {Danilov},\ and\ \citenamefont {Tzalenchuk}}]{Graaf2020}%
  \BibitemOpen
  \bibfield  {author} {\bibinfo {author} {\bibfnamefont {S.~E.}\ \bibnamefont {de~Graaf}}, \bibinfo {author} {\bibfnamefont {L.}~\bibnamefont {Faoro}}, \bibinfo {author} {\bibfnamefont {L.~B.}\ \bibnamefont {Ioffe}}, \bibinfo {author} {\bibfnamefont {S.}~\bibnamefont {Mahashabde}}, \bibinfo {author} {\bibfnamefont {J.~J.}\ \bibnamefont {Burnett}}, \bibinfo {author} {\bibfnamefont {T.}~\bibnamefont {Lindström}}, \bibinfo {author} {\bibfnamefont {S.~E.}\ \bibnamefont {Kubatkin}}, \bibinfo {author} {\bibfnamefont {A.~V.}\ \bibnamefont {Danilov}},\ and\ \bibinfo {author} {\bibfnamefont {A.~Y.}\ \bibnamefont {Tzalenchuk}},\ }\bibfield  {title} {\bibinfo {title} {Two-level systems in superconducting quantum devices due to trapped quasiparticles},\ }\href {https://doi.org/10.1126/sciadv.abc5055} {\bibfield  {journal} {\bibinfo  {journal} {Science Advances}\ }\textbf {\bibinfo {volume} {6}},\ \bibinfo {pages} {eabc5055} (\bibinfo {year} {2020})}\BibitemShut {NoStop}%
\bibitem [{\citenamefont {Miao}\ and\ \citenamefont {Gupta}(2009)}]{Miao2009}%
  \BibitemOpen
  \bibfield  {author} {\bibinfo {author} {\bibfnamefont {G.-X.}\ \bibnamefont {Miao}}\ and\ \bibinfo {author} {\bibfnamefont {A.}~\bibnamefont {Gupta}},\ }\bibinfo {title} {Growth and properties of epitaxial chromium dioxide (cro2) thin films and heterostructures},\ in\ \href {https://doi.org/10.1007/978-0-387-85600-1_17} {\emph {\bibinfo {booktitle} {Nanoscale Magnetic Materials and Applications}}},\ \bibinfo {editor} {edited by\ \bibinfo {editor} {\bibfnamefont {J.~P.}\ \bibnamefont {Liu}}, \bibinfo {editor} {\bibfnamefont {E.}~\bibnamefont {Fullerton}}, \bibinfo {editor} {\bibfnamefont {O.}~\bibnamefont {Gutfleisch}},\ and\ \bibinfo {editor} {\bibfnamefont {D.}~\bibnamefont {Sellmyer}}}\ (\bibinfo  {publisher} {Springer US},\ \bibinfo {address} {Boston, MA},\ \bibinfo {year} {2009})\ pp.\ \bibinfo {pages} {511--536}\BibitemShut {NoStop}%
\bibitem [{\citenamefont {Stephenson}\ and\ \citenamefont {Roth}(1971)}]{Stephenson1971May}%
  \BibitemOpen
  \bibfield  {author} {\bibinfo {author} {\bibfnamefont {N.~C.}\ \bibnamefont {Stephenson}}\ and\ \bibinfo {author} {\bibfnamefont {R.~S.}\ \bibnamefont {Roth}},\ }\bibfield  {title} {\bibinfo {title} {{Structural systematics in the binary system Ta2O5{\textendash}WO3. V. The structure of the low-temperature form of tantalum oxide L-Ta2O5}},\ }\href {https://doi.org/10.1107/S056774087100342X} {\bibfield  {journal} {\bibinfo  {journal} {Acta Crystallogr. B}\ }\textbf {\bibinfo {volume} {27}},\ \bibinfo {pages} {1037} (\bibinfo {year} {1971})}\BibitemShut {NoStop}%
\bibitem [{\citenamefont {Zagorac}\ \emph {et~al.}(2019)\citenamefont {Zagorac}, \citenamefont {M{\ifmmode\ddot{u}\else\"{u}\fi}ller}, \citenamefont {Ruehl}, \citenamefont {Zagorac},\ and\ \citenamefont {Rehme}}]{Zagorac2019Oct}%
  \BibitemOpen
  \bibfield  {author} {\bibinfo {author} {\bibfnamefont {D.}~\bibnamefont {Zagorac}}, \bibinfo {author} {\bibfnamefont {H.}~\bibnamefont {M{\ifmmode\ddot{u}\else\"{u}\fi}ller}}, \bibinfo {author} {\bibfnamefont {S.}~\bibnamefont {Ruehl}}, \bibinfo {author} {\bibfnamefont {J.}~\bibnamefont {Zagorac}},\ and\ \bibinfo {author} {\bibfnamefont {S.}~\bibnamefont {Rehme}},\ }\bibfield  {title} {\bibinfo {title} {{Recent developments in the Inorganic Crystal Structure Database: theoretical crystal structure data and related features}},\ }\href {https://doi.org/10.1107/S160057671900997X} {\bibfield  {journal} {\bibinfo  {journal} {J. Appl. Crystallogr.}\ }\textbf {\bibinfo {volume} {52}},\ \bibinfo {pages} {918} (\bibinfo {year} {2019})}\BibitemShut {NoStop}%
\end{thebibliography}%

\appendix
\section{$\mu$SR stopping profile details} \label{sec:appendix_a}

\begin{figure}[!htb]
\includegraphics[width=0.48\textwidth]{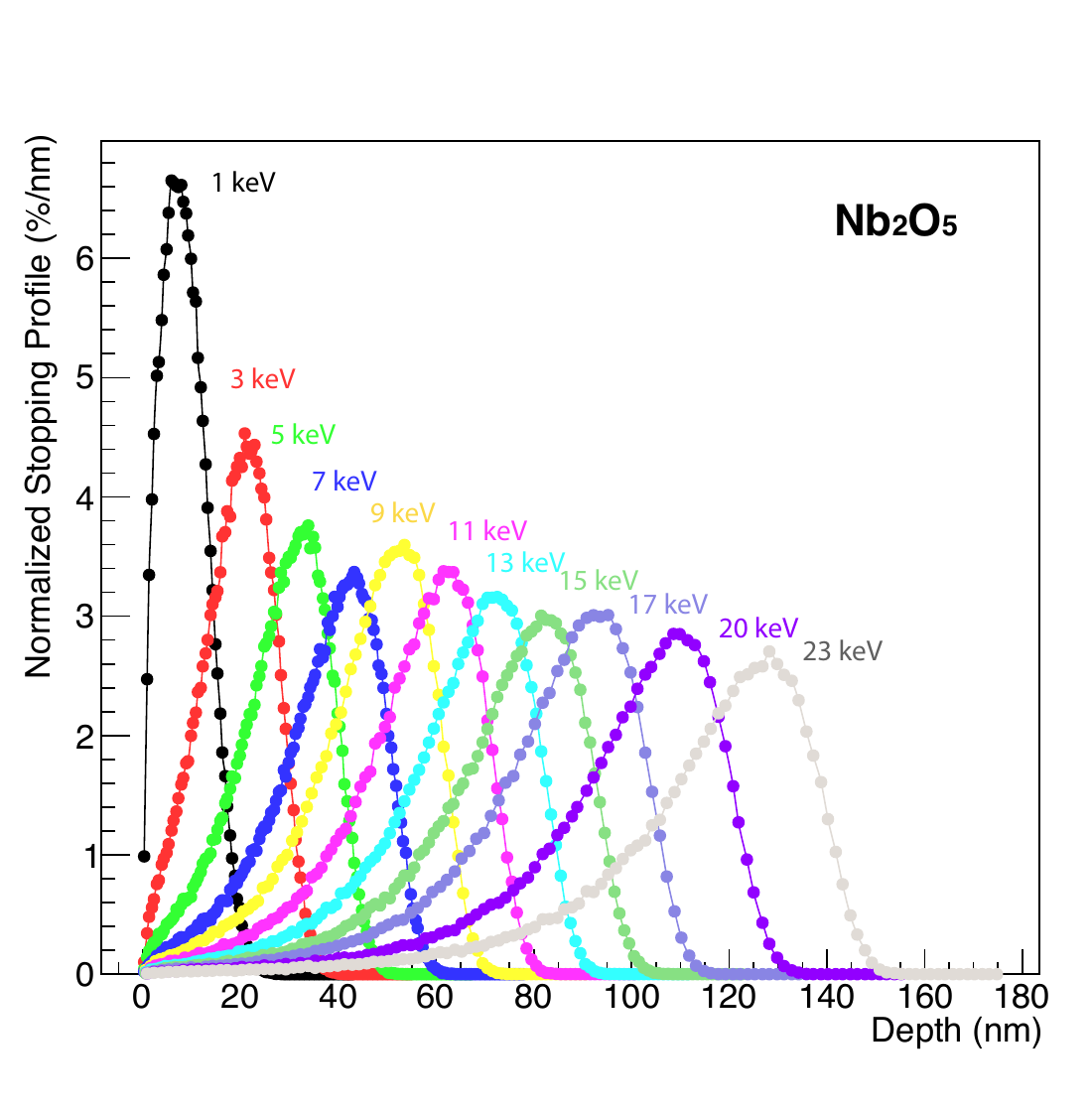}
\caption{\label{fig:profile_Nb2O5} Simulated muon stopping profile for Nb$_2$O$_5$ assuming a mass density of \SI{5.2}{\gram\per\cm^3}.}
\end{figure}

\begin{figure}[!htb]
\includegraphics[width=0.48\textwidth]{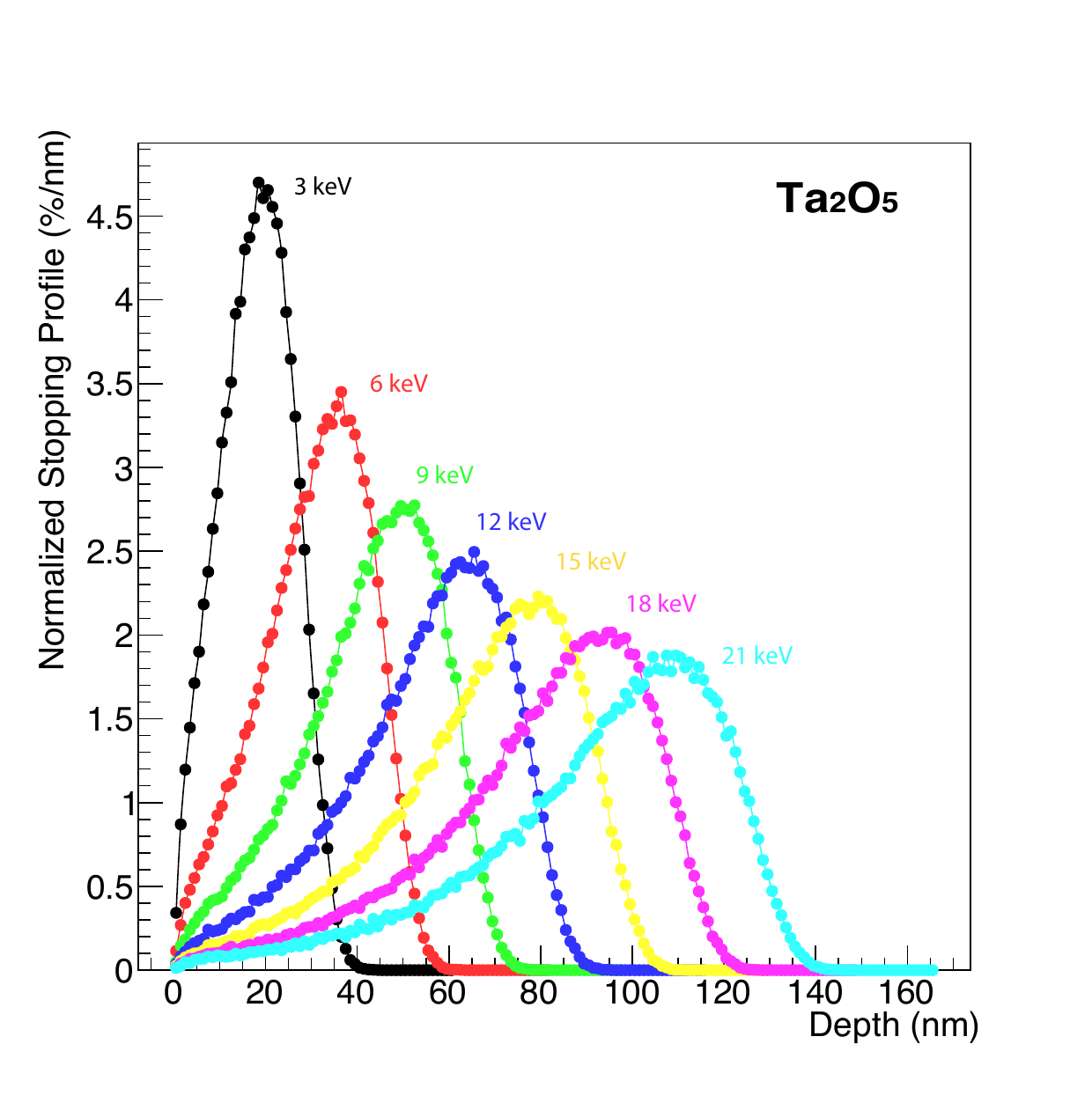}
\caption{\label{fig:profile_Ta2O5} Simulated muon stopping profile for Ta$_2$O$_5$ assuming a mass density of \SI{8.2}{\gram\per\cm^3}.}
\end{figure}

Calculated muon stopping profiles for Nb$_2$O$_5$ are shown in Fig.~\ref{fig:profile_Nb2O5} and for Ta$_2$O$_5$ in Fig.~\ref{fig:profile_Ta2O5} for the relevant muon beam energies.

\section{Materials characterization} \label{sec:appendix_b}
\subsection{\it{XPS studies}}

\begin{figure}[!htb]
\includegraphics[width=0.5\textwidth]{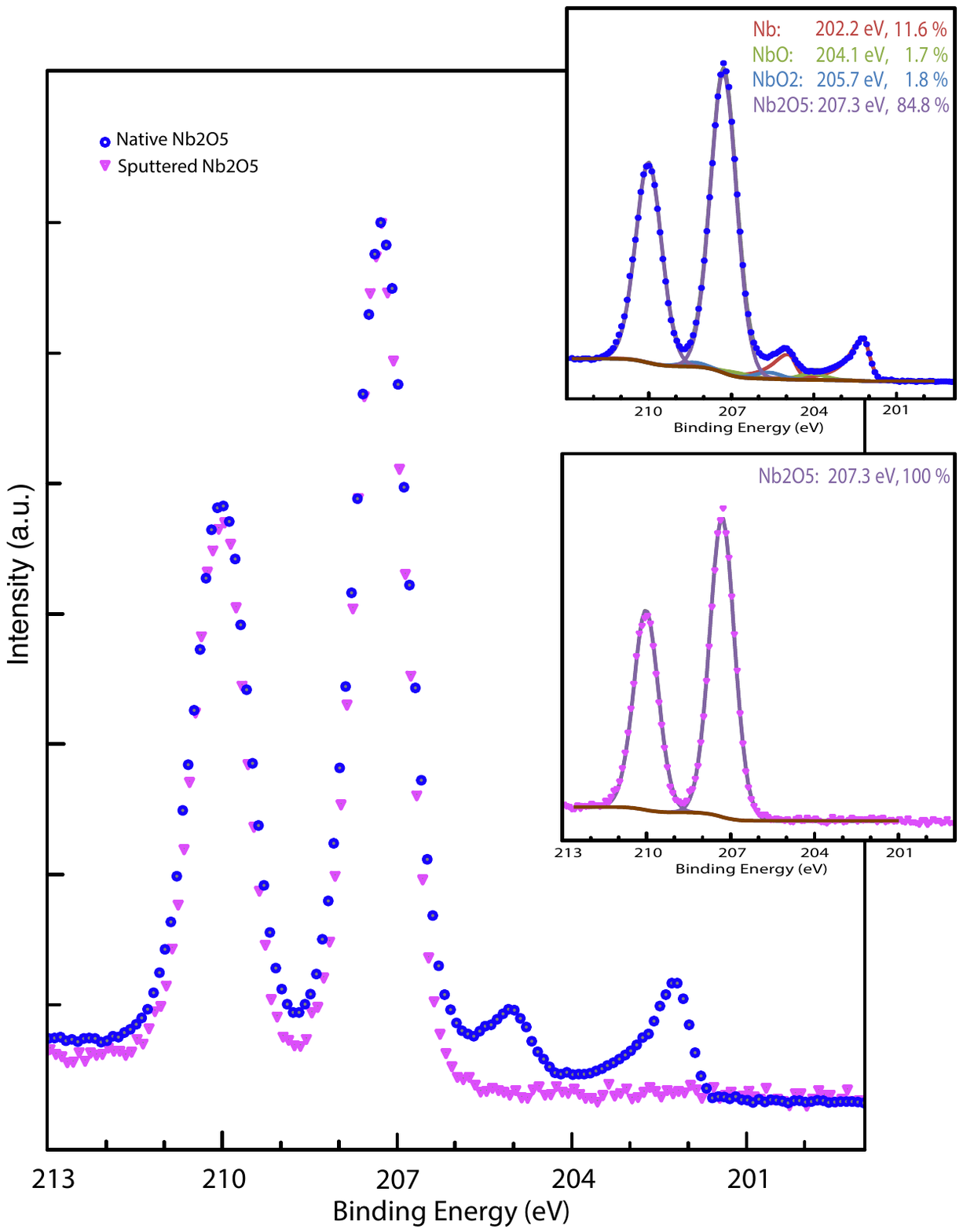}% Here is how to import EPS art
\caption{\label{fig:Nb3d} XPS Nb, Nb$_2$O$_5$ films. insets are fitted data for native surface Nb oxide and grown sputtered pentoxide film. }
\end{figure}

\begin{figure}[!htb]
\includegraphics[width=0.5\textwidth]{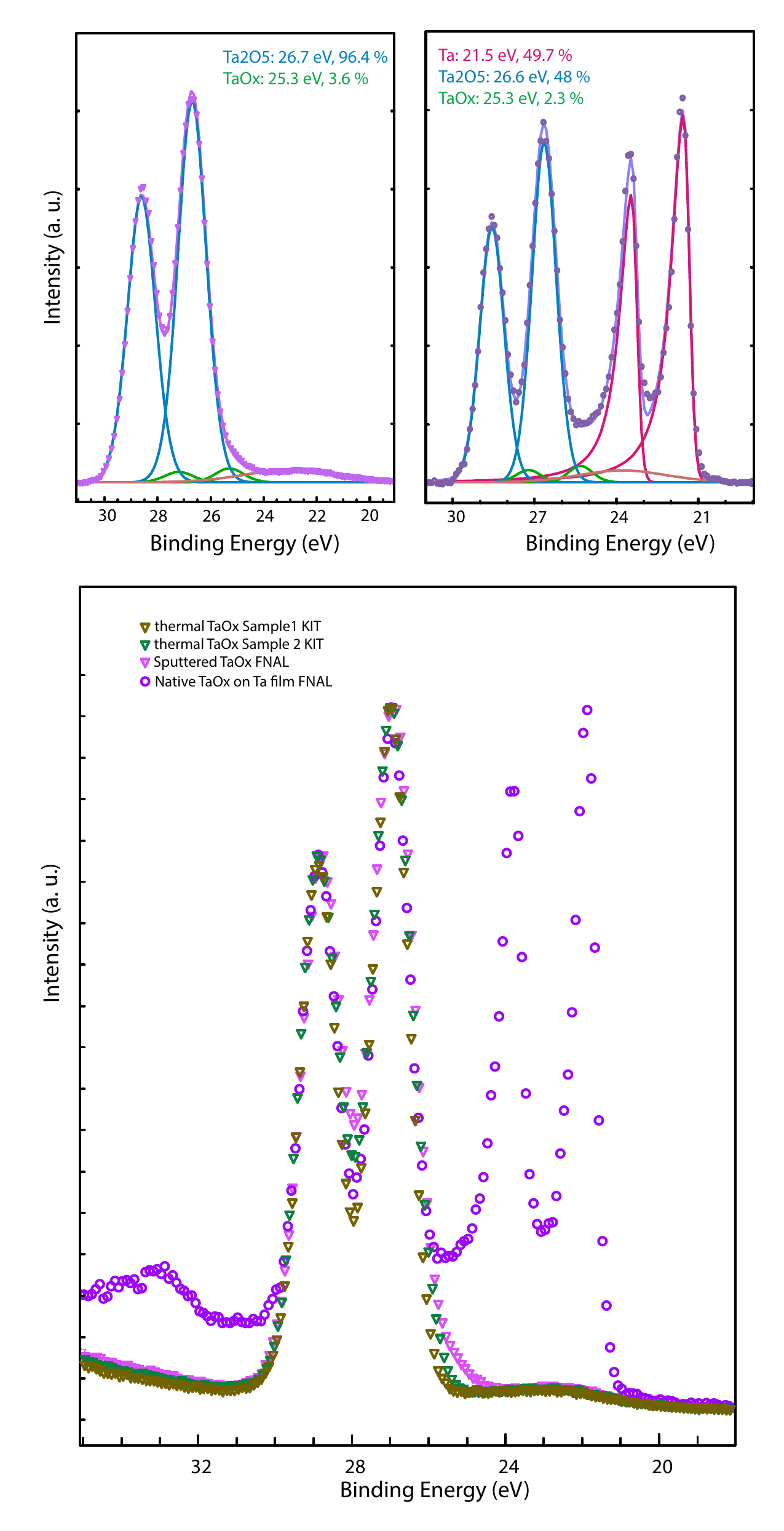}% Here is how to import EPS art
\caption{\label{fig:Ta4f} XPS Ta, Ta$_2$O$_5$ films. Upper panels are fitted data for native surface oxide and grown sputtered film. }
\end{figure}

X-ray photoelectron spectra were collected using a SPECS FlexMod-FlexPS, spot size  for FlexPS is 2~mm.  XPS analysis was performed using the Casa XPS. The background signal arising from inelastically scattered electrons was removed using a Shirley baseline model and the peaks were fit using a Gaussian-Lorentzian product.

X-ray photoelectron spectroscopy (XPS) was used to examine the chemical composition of these films as seen in Fig.~\ref{fig:Nb3d} and in Fig.~\ref{fig:Ta4f}. From the Nb 3d spectra, we observe that a majority of the Nb atoms in the oxide layer have a charge state of 5+, which corresponds to Nb$_2$O$_5$. Nonetheless, through comparison with similar measurements performed at the surface of a Nb film, we observe that the grown Nb$_2$O$_5$ film is chemically very similar to the native Nb$_2$O$_5$ that forms at the surface of Nb. The Nb$^0$ charge state peak results from photoelectrons that are generated in the Nb film that lies beneath the surface oxide in the sample. 
Since the XPS data gives only information about the topmost 7~nm of the film, we checked a depth profile using time of flight secondary ion mass spectrometry (ToF-SIMS). The results suggest that the stoichiometry of the Nb$_2$O$_5$ thin film sample remains fairly constant.

We performed the same XPS measurements for samples with sputtered and thermal Ta$_2$O$_5$ films as well as the native oxide on a Ta film for comparison. The resutls are shown in Fig.~\ref{fig:Ta4f}. The metallic tantalum 4f5/2 and 4f7/2 peaks are more pronounced for Ta film due to the thinner surface oxide layer when comparing to Nb. The Ta film data also show a loss feature around $33$~eV related to metallic Ta. The fit data suggest a minor presence of tantalum suboxides, which is very similar in the grown oxide films and the native surface oxide. 

The $\mu$SR results discussed in the manuscript cover solely oxygen-deficient pentoxides
measured under ultra-high vacuum conditions of 10$^{-8}$ mbar, excluding potential oxygen or hydrogen precipitation that could occur when cooling the samples.

\subsection{\it{Structure of Ta$_2$O$_5$ films}}

\begin{figure*}[!htb]
\includegraphics[width=\textwidth]{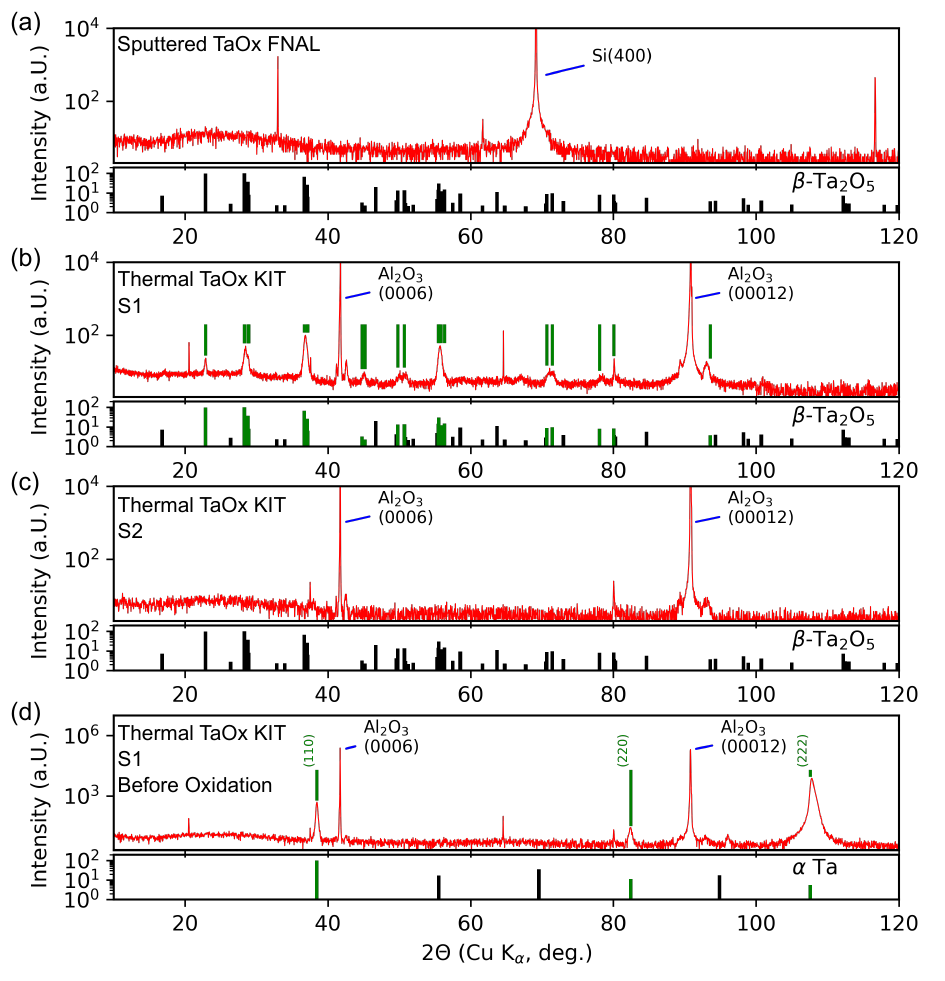}%
\caption{\label{fig:TaOx_XRD} High-resolution X-Ray diffraction analysis for Ta$_2$O$_5$ films. 
(a) X-Ray diffractogram of sputtered Ta$_2$O$_5$ on silicon providing evidence for the absence of crystallinity.  For reference the powder crystal diffraction pattern for $\beta-$Ta$_2$O$_5$ ($L-$Ta$_2$O$_5$) is shown in the plot below the diffractogram (ICSD collection code 9112).
(b) X-Ray diffractogram of thermal Ta$_2$O$_5$ film S1 on c-plane sapphire. Several diffraction peaks are visible, demonstrating that the thermally oxidized film is at least partially in a polycrystalline state. We could associate all peaks (marked in green) that are not related to the sapphire substrate with the $\beta-$Ta$_2$O$_5$ shown in the plot below it.
(c) X-Ray diffractogram of thermal Ta$_2$O$_5$ film S2 on c-plane sapphire. The absence of any other peaks confirms that the thermal Ta$_2$O$_5$ film on sample S2 was amorphous or contained crystallites too small to be detected by XRD. This we attribute to the smaller film thickness compared to sample S1. 
(d) X-Ray diffractogram of film S1 on c-plane sapphire before thermal oxidation. It can be seen that the Ta film on sample S1 was strongly textured and predominantly (111) oriented, with a minor contribution from (110) oriented crystallites.}
\end{figure*}

The structure of the Ta$_2$O$_5$ films are further characterized in Fig.~\ref{fig:TaOx_XRD} by
X-ray diffraction. We use a Bruker high-resolution X-ray diffraction system in reflection, equipped with a
$(022)$ Ge monochromator for the characteristic K$_{alpha}$ line of a Cu X-ray source.

The sputtered Ta$_2$O$_5$ sample does not exhibit any diffraction peaks originating from the film, which suggests that it is amorphous -- see Fig.~\ref{fig:TaOx_XRD}~(a).  The most prominent peak is the (400) silicon substrate peak. The sharp peaks at $2\Theta=33.0^{o},~ 116.5^{o}$ are associated with basis-forbidden (200) and (600) reflections from the silicon substrate that are visible due to multiple diffraction (Umweganregung)\cite{Zaumseil2015Mar}. The peak at $2\Theta=61.7^{o}$ is due to (400) reflection of some remaining Cu $K_{\beta}$ radiation, that was not fully suppressed by the monochromator. Some diffuse intensity can be seen in the range $2\Theta=15^{o}-30^{o}$ likely associated with the amorphous Ta$_2$O$_5$ film.

The diffractogram of the thermally oxidized Ta$_2$O$_5$ sample S1 - see  Fig.~\ref{fig:TaOx_XRD}(b) -shows that it is at least partially polychrystalline. We could match the observed diffraction peaks with a pattern derived from a beta-Ta2O5 structure (ICSD-9112, ICSD-release~2025.1)~\cite{Zagorac2019Oct, Stephenson1971May}. 

The diffractogram of the thermally oxidized Ta$_2$O$_5$ sample S2 is shown in Fig.~\ref{fig:TaOx_XRD}~(c). It does not show any diffraction peaks associated with the film, likely due to the lower film thickness.
The minor sharp peaks at $2\Theta=37.5^{o}$ and $2\Theta=80.0^{o}$ are due to some remaining Cu $K_{beta}$ radiation, that was not fully suppressed by the monochromator. The satellite peaks of the Sapphire substrate peaks are very likely also related to the substrate.

 For reference we also provide the X-ray data of sample S1 before oxidation (see Fig.~\ref{fig:TaOx_XRD}~(d)). Comparing the Ta film peaks to the diffractograms of the thermally oxidized films, shows these completely absent after oxidation.

\begin{figure*}[!htb]
\includegraphics[width=0.7\textwidth]{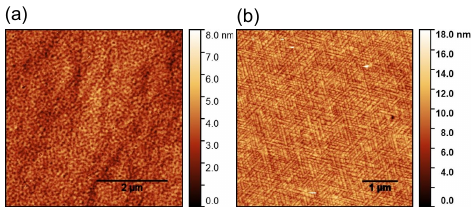}%
\caption{\label{fig:ThermalTaOx_AFM} Atomic-force microscopy (AFM) of thermally oxidized sample before (a) and after oxidation (b).   }
\end{figure*}

Since thermal Ta$_2$O$_5$ sample S2 was shown to be at least partially poly-crystalline we also investigated the change in the film due to the oxidation process with atomic-force microscopy as shown in Fig.~\ref{fig:ThermalTaOx_AFM}. The surface corrugation increases due to oxidation. After oxidation the morphology clearly follows the hexagonal symmetry of the c-plane sapphire surface.

\subsection{\it{Sputtered Ta$_2$O$_5$ Squid Magnetometry}}

\begin{figure*}[!htb]
\includegraphics[width=0.8\textwidth]{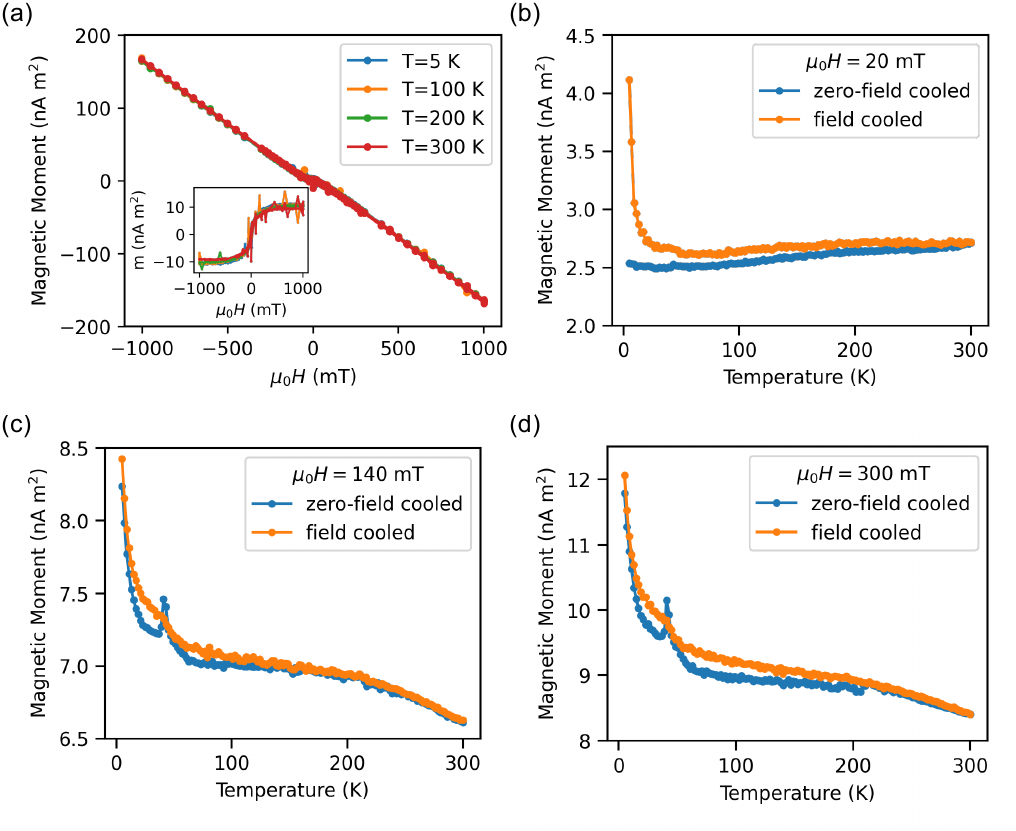}%
\caption{\label{fig:squid} SQUID Magnetometry of sputtered Ta$_2$O$_5$ thin film on silicon substrate. (a) The negative slope of the magnetic moment $m$ as a function of applied field $H$ indicates that the sample's magnetization is predominantly diamagnetic. In the inset we have subtracted a constant slope of $-17.5 \times 10^{-5}$~ emu / T. The same correction has been applied to the zero-field cooled/field-cooled temperature dependent measurements for fields (b) $B_{\rm ext}=20$~mT, (c) $140$~mT and (d) $300$~mT. }
\end{figure*}

Since the sputtered Ta$_2$O$_5$ films exhibited a large magnetic volume fraction in transverse field muSR we characterized the sample in a SQUID magnetometer (Quantum Design). In particular the aim was to investigate if the magnetic transition at around $225$~K can be detected in the macroscopic magnetic moment of the film. The results of a field dependent measurement and temperature dependent zero-field-cooled/field-cooled measurements are presented in Fig.~\ref{fig:squid}. The field dependence of the sample's magnetic moment is weakly diamagnetic and temperature independent, as expected from the silicon substrate. The sample mass was $15.335$~mg.  On subtracting this diamagnetic component, a small soft ferromagnetic response with Curie temperature well above $300$~K remains. This is likely due to residual contaminants not related to the film, as the contribution could be reduced significantly from previous measurements by solvent cleaning the sample. The temperature dependent measurements are overall dominated by the residual contaminant.  There is an increasing paramagnetic moment below $10$~K most likely due to paramagnetic impurities in the substrate. The small feature around $50$~K is due to some remaining oxygen impurity in the exchange gas of the squid magnetometer. We note a small irreversible component below $225$~K for $B_{\rm ext}=300$~mT, which could be potentially related to the transition observed in $\mu$SR.

\end{document}